\documentclass[12pt]{article}

\usepackage{arxiv}

\usepackage[utf8]{inputenc} 
\usepackage[T1]{fontenc}    
\usepackage{hyperref}       
\usepackage{url}            
\usepackage{booktabs}      
\usepackage{amsfonts}      
\usepackage{nicefrac}       
\usepackage{microtype}      
\usepackage{lipsum}
\usepackage{graphicx}
\usepackage{subcaption}
\usepackage[export]{adjustbox}
\usepackage{chngcntr}
\graphicspath{ {./images/} }
\usepackage{natbib}
\usepackage{float}
\usepackage{longtable}
\usepackage{setspace}
\usepackage{adjustbox}
\usepackage{rotating}
\usepackage{enumitem}
\setcitestyle{round}
\title{Adopting AI: How Familiarity Breeds Both Trust and Contempt}

\author{ 
    Michael C. Horowitz \\
    University of Pennsylvania\\
   \AND
   Lauren Kahn\\
   Council on Foreign Relations\\
   \AND
   Julia Macdonald\\
   University of Denver\\
   \AND
   Jacquelyn Schneider\\
   Stanford University\\
}
       
\begin{document}

\maketitle

{\centering {\large Accepted for publication at AI \& Society: Knowledge, Culture, and Communication}}

\clearpage

\begin{abstract}
Despite pronouncements about the inevitable diffusion of artificial intelligence and autonomous technologies, in practice it is human behavior, not technology in a vacuum, that dictates how technology seeps into--- and changes ---societies. In order to better understand how human preferences shape technological adoption and the spread of AI-enabled autonomous technologies, we look at representative adult samples of US public opinion in 2018 and 2020 on the use of four types of autonomous technologies: vehicles, surgery, weapons, and cyber defense. By focusing on these four diverse uses of AI-enabled autonomy that span transportation, medicine, and national security, we exploit the inherent variation between these AI-enabled autonomous use cases. We find that those with familiarity and expertise with AI and similar technologies were more likely to support all of the autonomous applications we tested (except weapons) than those with a limited understanding of the technology. Individuals that had already delegated the act of driving by using ride-share apps were also more positive about autonomous vehicles. However, familiarity cut both ways; individuals are also less likely to support AI-enabled technologies when applied directly to their life, especially if technology automates tasks they are already familiar with operating. Finally, opposition to AI-enabled military applications has slightly increased over time.
\end{abstract}


\doublespacing
\clearpage

\section{Introduction}

In 2021, Alphabet CEO Sundar Pichai declared that artificial intelligence (AI) was “the most profound technology humanity will ever work on,”--a larger driver of societal change than “fire or electricity or the internet” \citep{Steiner2021}. Pichai’s pronouncement is not unique. Future trend-spotters in transportation, healthcare, and warfare all foresee an autonomous future with autonomous cars, robotic surgeons, and drones changing the way humans interact,compete, and survive \citep{Fryer-Biggs2019,MedicalFuturist2021,Gupta2021}.
    
However, while artificial intelligence and the way it enables autonomous applications may be diffusing across societies, technology does not diffuse without human intervention. It is human behavior, not necessarily technology in a vacuum, that dictates the vagaries of how technology seeps into and changes societies \citep{MacKenzie1993,Slayton2013,Herrera2006,Jasanoff2004,LeeTrimiKim2013}. Whether it is the adoption of artificial intelligence in consumer goods, infrastructure, or national security—quite often it is the consumer, the citizen, the taxpayer, and the soldier who dictates the use and growth of new technologies. As Hall and Khan argue about the adoption of new technologies, “it is diffusion rather than invention or innovation” \citep{HallKhan2004} that ultimately determines the impact of technologies \citep{Horowitz2010}. If artificial intelligence is as ubiquitous or as profound as its proponents claim, then its reach across economies and societies makes for a fascinating phenomenon that impacts how individuals vote, economies and markets evolve, regimes govern, and when and why states go to war \citep{Levy2018,Horowitz2018,ZhangEtAl2008,HelbingEtAl2019,Bissell2018}.
    
To better understand technological adoption and the spread of AI-enabled autonomous technologies today, we look at representative adult samples of US public opinion in 2018 and 2020 on the use of four types of autonomous technologies: vehicles, surgery, weapons, and cyber defense. By focusing on these four diverse uses of AI-enabled autonomy that span transportation, medicine, and national security, we exploit the inherent variation between these AI-enabled autonomous use cases. This includes both uses of AI with greater salience for the public (self-driving vehicles), potential applications relevant to individual well-being (robotic surgery), and both offensive and defensive military applications (autonomous weapon systems and cyber defense). 

We theorize that support for AI-enabled autonomous technologies depends in part on familiarity and trust, even across use cases \citep{schepman2020}. We further theorize that there are delegation effects whereby people who have already made the decision to delegate specific tasks, such as driving, to other humans might be more supportive of delegating those same tasks to artificial intelligence technologies. Additionally, the variation between 2018 and 2020 provides a novel mechanism that allows us to examine directly how attitudes about AI adoption change over time, and what factors might drive these shifts.
    
We find that those with familiarity and expertise with AI and similar technologies were more likely to support all of the autonomous applications we tested (except weapons) than those with a limited understanding of the technology. We find support for the theorized delegation effect when it comes to autonomous vehicles, but less support when technology automated tasks with which individuals were not familiar operating. Finally, opposition to AI-enabled military applications slightly increased over time. \footnote {Additionally, though there are reasons to think that over time and with COVID-19 as an intervening factor between 2018 and 2020, there might be an increase in support for AI-enabled autonomy, we find little evidence that time and a pandemic made individuals more likely to adopt autonomous technologies \citep{ccescovid}.} Our findings suggest a complicated relationship between users and AI-enabled technologies where familiarity with AI may instill trust, but only up to a point. The old saying that "familiarity breeds contempt" could help explain why users are less likely to adopt automated versions of technologies with which they are already accustomed to operating without any AI interventions or accept machine intervention when accustomed to more direct human involvement.  
    
Below, we first introduce existing data on public support for AI-enabled autonomous systems. We then use the literature on technology diffusion to examine existing theories about the determinants of technological adoption. In doing so, we identify a series of hypotheses that we then test in our subsequent data section. We then turn to a discussion of demographic and intervening variables before concluding.

\section{Theory}

For the purposes of this paper, we define artificial intelligence as the capability for machines to conduct tasks once thought to require human intelligence \citep{russell2020}. Artificial intelligence methods like machine learning are one way to program autonomous systems or systems that operate with minimal or no human oversight. Past research finds that public attitudes toward AI-enabled autonomous systems tend to vary based on the technology's application. For example, support for higher-risk technologies, such as autonomous vehicles, has remained relatively sticky over time \citep{West2018}, with higher levels of support among young, high-income males within the tech field \citep{PayreEtAl2014,BansalEtAl2016,HulseEtAl2018}. Similarly, support for the development of AI technologies for use in warfare has remained relatively low at 30\% (though support increases to 45\% when adversaries develop similar weapons \citep{West2018}). More general questions about AI in the same survey reveal a conflicted public; when asked whether AI “is a good thing/bad thing for society,” 44\% of US adults said it was a good thing, while 47\% said it was a bad thing. 


Existing surveys and polls provide valuable snapshots of support for AI technologies and their change over time. However, what these surveys struggle to explain is what drives support for the adoption of AI-enabled autonomous systems in the first instance --- are changes in support due to simply a greater exposure to and awareness of the technology as it develops and becomes more prevalent or due to changes in how the technology can be used in new contexts and how much control an individual has over the system? However, the existing literature on technology adoption and diffusion suggests a series of hypotheses that drive the heart of this puzzle.

\subsection{Prior experience, familiarity, \& knowledge}

One of the primary factors that might influence support for the use of AI-enabled technology is familiarity with the technology. Existing research suggests that understanding the application of algorithms in the real world—and greater familiarity with autonomy in general—might lead to recognition of possibilities, and appreciation of technical limits \citep{YarbroughSmith2007,MarangunicGranic2015,Chau1996,KingHe2006}. For example, research in the medical field shows doctor familiarity with computing in the late 1990s and early 2000s made the adoption of computerized healthcare processes more likely \citep{AustinEtAl2006,LapinskyEtAl2004}. Similarly, research on the adoption of autonomous vehicles suggests those with careers in high-technology fields were more likely to support the development of fully autonomous vehicles \citep{BansalEtAl2016,PayreEtAl2014,MoodyEtAl2020} and that those that had more familiarity with autonomous vehicles directly were more likely to find them safe \citep{PenmetsaEtAl2019,SanbonmatsuEtAl2018}.

More broadly, behavioral psychology research illustrates the link between personal experience, familiarity, and support for technologies \citep{TaylorTodd1995}. Direct experience influences how people process information. When an individual believes they have experience with a concept or application, it makes them more empathetic to the concept or application, ultimately viewing it in a positive light \citep{FazioZannaCooper1978,FazioZanna1981}. Prior experience also makes it easier to rely on one's own judgment when making an assessment, rather than the opinions of others \citep{BurnkrantCousineau1975}. This is particularly true when considering information systems—exposure generates favorable attitudes towards future adoption \citep{HartwickBarki1994}. This is because personal experience and increased familiarity can generate a greater sense of knowledge about, and confidence in, the use of a given technology. Prior survey research of the general public shows that individuals are more comfortable adopting new technologies once they are familiar with them. 52\% of the public prefers using familiar brands and products, and only 35\% wants to try new technologies without additional evidence of effectiveness. 39\% describe themselves as preferring to wait until they hear about others’ experiences before trying something new themselves \citep{KennedyFunk2016}. It is also true when looking at specific research on artificial intelligence, which has shown that factors such as comfort with specific applications are often even stronger predictors of general attitudes towards AI than the perceived capability of the AI itself \citep{schepman2020}. Similarly, other attempts to test confidence in AI systems have found that different types of direct experience with AI (either positive or negative) have a significant impact on not only how humans approach using AI systems, and particularly, AI-enabled decision aids, but their self-confidence in completing a task, as well \citep{Chong2022}.

\noindent\textit{Hypothesis 1: Greater familiarity with AI, through knowledge and self-reported use, should lead to greater support for uses of AI.}

\subsection{Delegation}

But familiarity is not always enough to lead to adoption. We theorize that people's attitudes towards AI-enabled technologies are also determined by a variable that interacts with familiarity---whether individuals are already comfortable with delegating decision-making for the task. Many AI-enabled technologies require individuals to delegate some degree of decision-making power---whether selecting grocery produce, making smart banking choices, or driving around town. What makes an individual more or less willing to delegate decision-making to a machine? 

In general, previous research suggests people are more likely to delegate to AI-enabled technologies in situations where they have already delegated authority or control over the activity to another human or technology \citep{Milleretal2007}. When individuals have already ceded some decision-making powers---for example, to ridesharing app drivers---they have already made the decision to trust another agent, and therefore the decision to delegate to AI-enabled technologies should be easier than for those who have not. 

In contrast, when individuals currently conduct the task themselves, they may be less willing to delegate responsibility to a machine. Their familiarity and experience with operating the technology make them more distrustful of machine intervention. For example, research on the adoption of autonomous vehicles finds experienced drivers often question whether autonomous vehicles are safe enough to adopt \citep{KonigNeumayr2017}. They value the control, even though it involves a greater cognitive load for themselves \citep{Milleretal2007}. More abstractly, research on AI shows that despite a potential aversion to entrusting strategic decisions to algorithms \citep{LeyerSchneider2019}, delegation to an algorithm is easier if someone has already transferred control of a task in the first place, because using the algorithm only requires trusting the algorithm, not delegating the decision in the first place \citep{HeberSchneider2020}. Part of the logic here involves direct experience with the task, since “in general, any form of task delegation - whether to automation or other humans - must necessarily result in added unpredictability if it offloads tasks" \citep{Milleretal2007}.

We can evaluate delegation and support for AI-enabled autonomous systems in a few ways. First, we can measure whether the individual already delegates driving via the use of ridesharing apps. We would predict individuals who have delegated control to ridesharing apps are more likely to support autonomous vehicles than those who have not.

\noindent\textit{Hypothesis 2a: Those that used ridesharing apps prior to the pandemic should be more supportive of autonomous vehicles than those surveyed in the 2020 CCES.}

Second, individuals already delegate responsibility when it comes to undergoing surgery in hospitals. Even those who attempt to manage their own healthcare have to trust others when it comes to operations. Therefore, since people have already decided to delegate surgery to a doctor, the decision to trust an algorithm may be easier than when compared to trusting an algorithm with something they currently do themselves.

\noindent\textit{Hypothesis 2b: Support for autonomous surgery should be higher than support for autonomous vehicles.}


\subsection{Defense Applications: AI-Enabled Weapons vs. AI-Enabled Cyber Defense}

Autonomous vehicles and surgeries use artificial intelligence for tasks society sees as generally beneficial. Can the same theories of familiarity and delegation explain public support for AI-enabled weapons and cyber defense?  These are tough cases.  The public is certainly familiar with the idea of remotely-piloted aircraft and AI-enabled weapons. The \textit{Campaign to Stop Killer Robots} and public figures like Elon Musk have quickly raised public and elite awareness about the potential dangers surrounding highly autonomous weapon systems. Popular science fiction TV, movies, and books also make it easier to imagine the worst-case scenario possibilities of developing and using autonomous weapon systems. Previous surveys have found the US public and AI experts alike, to be wary about the use of autonomy and artificial intelligence within offensive military operations \citep{Horowitz2016, ipsos, zhang2021ethics}. 

However, even though the average American may be familiar with AI-enabled weaponry as presented in the media, they have little to no familiarity with operating or experiencing these technologies within their own lives. This makes AI-enabled weapons different than other technologies like autonomous cars or even autonomous surgeries--which normal Americans are more likely to experience or engage with in their day-to-day lives. When forming opinions about delegating tasks to AI-enabled weaponry, the public's perception of familiarity is tempered by their actual lack of experience using these technologies. Thus, unlike other uses of AI-enabled technology which should see a general increase in support based on familiarity with AI technologies, we don't expect this to occur with AI-enabled weapons.

Further, public support for the use of force, and their willingness to delegate to the military decisions about the use of force on the public's behalf, is complicated \citep{feavergelpi2011, jentleson1992pretty}. The American public is generally concerned about  civilian collateral damage and weapons are seen as more likely to harm civilians are more likely to meet with public disapproval \citep{walsh2015precision, schneider2016us}. Therefore, with AI-enabled weapons, we introduce a case with potential perceptions of high familiarity, but actual low familiarity and subsequent high concerns about delegation. 

This is a particularly interesting case to explore familiarity and delegation because it tests whether these concepts help explain technologies that the public will likely never be familiar nor experienced with using in their day-to-day lives. Unlike autonomous vehicles or even autonomous surgery, the average American already delegates their defense to the military to an extent, but in a much more extreme fashion given the average person does not have any control or direct recourse when it comes to national defense. Moreover, while this might initially suggest greater levels of support, public debate and concern over the ethics and safety of the introduction of AI into military contexts, such as the \textit{Campaign to Stop Killer Robots}, likely overwhelms any potential delegation effect.  Given limited familiarity with non-sensationalized uses of the technology, and a general reluctance to delegate responsibility to machines in war that are associated with high potential for accidents or collateral damage, we expect that, in general, support for AI in weapons systems will be lower than AI use in more public utility functions such as autonomous vehicles or AI-enabled surgeries. 

\noindent\textit{Hypothesis 3: Support for AI will be lowest when applied to autonomous weapon systems}.



\subsection{Policy Support vs. Personal Use}

While the baseline questions about AI-enabled autonomous system adoption focus on support for the use of AI in a particular arena, such as autonomous vehicles, support for use of AI as a matter of public policy may differ from personal beliefs about or willingness to use. People process information differently when it involves their own experiences or potential experiences, especially when it involves risk to themselves. Specifically, support for use of AI as a matter of public policy for areas such as autonomous vehicles may be higher than the willingness of the same respondents to ride in an autonomous vehicle. For instance, in a 2014 study of autonomous vehicle adoption, researchers found a significant disparity between participants' general support for fully autonomous vehicles and actual willingness to buy these vehicles \citep{PayreEtAl2014}, with trust and risk playing the most important role in distinguishing between general support and willingness to pay for the technology \citep{LiuEtAl2019}. Essentially, as people have to shift from thinking about adoption from a societal perspective---a public policy judgment---to thinking about adoption from an individual perspective---their use of AI in a particular area---safety and reliability concerns are likely to grow, leading to greater opposition.

This concept of a support-use gap is reinforced by existing research on how confidence and trust influence human-machine relationships \citep{MacdonaldSchneider2019}. Trust inherently involves a degree of uncertainty since it involves having to rely on another. It is “the willingness to make oneself vulnerable to another based on a judgment of similarity of intentions or values” \citep{SiegristEtAl2005}. Trust is important in helping to facilitate choices in situations characterized by uncertainty, vulnerability, and perceived risk, where the “motives, intentions, and prospective actions of others” are unknown \citep{Kramer1999,JosangPresti2004}. 


AI is a newer technology that elicits a degree of public concern. Studies of American public opinion show that, in general, only about 18\% of those surveyed are "more excited than concerned" about "the increased use of AI in daily life."\citep{Pew2022} In terms of public policy preferences, we would expect this to translate into stronger support for AI adoption for society overall, and for individuals to be less supportive of personally using AI, and that these trends will be magnified for those applications that pose higher levels of risk. 

\noindent\textit{Hypothesis 4: Support for broad AI adoption will be higher, on average, than a willingness to personally use AI, across all AI applications}

\section{Research Design}

We test our hypotheses about support for AI-enabled autonomous systems by evaluating questions in the 2018 and 2020 waves of the Cooperative Congressional Election Study (CCES), now called the Cooperative Election Study (CES) \citep{Schaffner2019, Schaffner2021}. All data will be publicly available at a Dataverse page upon article publication.

Both the 2018 and 2020 samples are representative of the US adult public, based on the CCES/CES methodology \citep{Schaffner2019, Schaffner2021}. The 2018 survey was fielded on 1000 individuals in two phases---before and after the November 2018 general elections in the United States, and the 2020 survey was also fielded on 1000 individuals in two phases---before and after the November 2020 general elections in the United States. A module in the 2018 CCES featured questions about attitudes surrounding the adoption of autonomous systems and artificial intelligence across the areas described above. We then included the same questions in the 2020 CES, but with additional covariates to test the hypotheses above. The study was preregistered using Open Science.\footnote{Pre-registered on Open Science at https://osf.io/854dq/. The hypotheses above are consistent with those that were preregistered.}

As there was not a substantial change in AI-enabled autonomous systems that would be salient to the general public, there should not be a technology-based driver of a shift in attitudes. We can further control for the impact of demographic factors and partisanship in regression models (see the appendix). We present the results below without team sample weights, but we show in the appendix the results are identical when adding team weights designed to make the sample even more representative.

The dependent variables come from four sets of questions asking respondents about their support for the adoption of AI-enabled autonomous systems: autonomous vehicles, autonomous surgery, autonomous cyber defense, and autonomous weapon systems. Each support question is measured on a four-point scale, where 1 represents very unsupportive and 4 represents very supportive. Full details on the coding of each item are available in the appendix. We describe our key independent variables of interest and control variables below. All come from the CCES/CES data unless explicitly described otherwise. We include a number of individual difference variables, such as age and level of education, in the table here because we use them as control variables in some of the regression models below, even though we do not theorize about them in the hypotheses above.

\setlist{nolistsep}
\begin{itemize}[noitemsep,wide, nosep, labelindent = 0pt, topsep = 1ex]
    \item Sex (1 if Female, 0 if Male)
    \item Age (Count)
    \item Race (1 if a respondent identified as White, 0 otherwise)
    \item Prior Military Service (1 if yes, 0 otherwise)
    \item Level of education (1-6, where 1 = did not complete high school and 6 = graduate degree)
    \item Partisanship (1-7, where 1 = strong Democrat and 7 = strong Republican)
    \item Use of Ridesharing Apps (1 if respondent has used ridesharing apps before COVID-19 pandemic, 0 otherwise.\footnote{Asked in 2020 only.})
    \item Drive (1 if respondent has a driver's license and 0 otherwise.\footnote{Asked in 2020 only.})
    \item Urbanization (1-4, where 1 = living in a city and 4 = living in a rural area)
    \item Self-reported level of prior experience with AI (0-5 scale where 0 is lowest and 5 is highest) (2020 version)
    \item Self-reported level of prior experience with AI (0-2 scale where 0 is lowest and 2 is highest) (2018 version)
\end{itemize}
 
The measure of prior knowledge and experience with AI contains three parts in the 2020 survey, and we can decompose them to see what kinds of prior self-reported experience actually lead to more positive attitudes about AI-enabled autonomous systems. The first part of the measure is a question that asks respondents if they use AI at home, at work, both, or neither. This measures whether people have exposure to algorithms in their daily lives. The second part of the measure is a question that asks people whether they think of themselves as using algorithms when using services that make media suggestions based on user history, like the Netflix selection algorithm. The third part of the measure is two questions testing respondent knowledge about artificial intelligence methods such as machine learning. We aggregate these into an index. The index score is 0 if someone answers no to the home/work question and the music/movies question, and gets both of the knowledge questions wrong. The index score is 5 if someone answers yes to everything and gets the knowledge questions correctly. The 2018 survey only asked the first question, so the distribution is very different, running from 0 to 2. We do not compare the impact of AI knowledge from 2018 to 2020 for this reason. Combined, for the 2020 survey, however, we can generate an index of self-reported prior use and knowledge that should lead to greater support for AI adoption, following the literature on AI support indices \citep{schepman2020, Parasuraman2014}.

Table \ref{SummaryStatsCombined} highlights the distribution of our key demographic variables across the 2018 and 2020 CCES.

\begin{table}[!htbp] \centering 
  \caption{Summary Statistics COMBINED} 
\begin{tabular}{@{\extracolsep{5pt}}lccccc} 
\\[-1.8ex]\hline\hline \\[-1.8ex] 
Variable & \multicolumn{1}{c}{N} & \multicolumn{1}{c}{Mean} & \multicolumn{1}{c}{St. Dev.} & \multicolumn{1}{c}{Min} & \multicolumn{1}{c}{Max} \\ 
\hline \\[-1.8ex] 
Gender (2018)             &     1000.00&        0.58&        0.49&        0.00&        1.00\\
Gender (2020)              &     1000.00&        0.56&        0.50&        0.00&        1.00\\
Age (2018)                 &     1000.00&       49.03&       17.75&       19.00&       96.00\\
Age (2020)                 &     1000.00&       49.29&       17.64&       19.00&       89.00\\
White (2018)               &     1000.00&        0.74&        0.44&        0.00&        1.00\\
White (2020)               &     1000.00&        0.72&        0.45&        0.00&        1.00\\
Level of Education (2018)  &     1000.00&        3.63&        1.54&        1.00&        6.00\\
Level of Education (2020)  &     1000.00&        3.61&        1.49&        1.00&        6.00\\
Family Income (2018)       &      901.00&        6.28&        3.34&        1.00&       16.00\\
Family Income (2020)       &      913.00&        6.32&        3.45&        1.00&       16.00\\
Partisanship: 1 = Dem, 7 = GOP (2018) &      960.00&        3.72&        2.24&        1.00&        7.00\\
Partisanship: 1 = Dem, 7 = GOP (2020)&      951.00&        3.48&        2.18&        1.00&        7.00\\
Prior AI Knowledge (2018 version)  &     1000.00&        0.47&        0.71&        0.00&        2.00\\
Prior AI Knowledge (2020 version)  &     1000.00&        1.26&        1.13&        0.00&        5.00\\
Current or Prior Military Service (2018) &     1000.00&        0.11&        0.31&        0.00&        1.00\\
Current or Prior Military Service (2020)&     1000.00&        0.11&        0.31&        0.00&        1.00\\
Urban Area (2018)          &      992.00&        2.24&        1.06&        1.00&        4.00\\
Urban Area (2020)          &      995.00&        2.20&        1.05&        1.00&        4.00\\
Pre-COVID Ridesharing Use (2020)&      933.00&        1.60&        0.90&        1.00&        5.00\\
Drivers License (2020)     &      997.00&        0.88&        0.33&        0.00&        1.00\\
COVID-19 Death Family/Friends (2020)&     1000.00&        0.16&        0.37&        0.00&        1.00\\
\hline\hline \\[-1.8ex]
\end{tabular} 
  \label{SummaryStatsCombined} 
\end{table}

\section{Results}

We start by assessing average levels of support for AI-enabled autonomous systems. Figure \ref{fig:AllMeans} below illustrates the mean level of support, concern over safety, and willingness to use these applications in 2018 and 2020 amongst the US adult public. Overall, support slightly decreased for most AI-enabled autonomous systems, though the results are broadly stable---surgery dropped from $2.482$ to $2.374$, weapons systems from $2.199$ to $2.03$, and cyber defense from $2.568$ to $2.362$. Support for autonomous vehicles increased slightly, from $2.354$ to $2.411$, but the change was not statistically significant. There is substantial variation in the magnitude of support depending on the application of the technology, with cyber defense, surgery, and vehicles, generally receiving more support than autonomous weapon systems.

Responses to the 'Willingness to use' question followed a similar pattern, with respondents being less likely to opt into autonomous surgery in 2020 ($2.319$) than in 2018 ($2.078 $), or support the use of autonomous weapon systems (decreased from $2.373$ to $2.313$) or cyber defense systems (decreased from $2.457$ to $2.227$) in high-importance missions. Willingness to ride in an autonomous vehicle, however, increased by a small margin from $2.14$ to $2.186$.

Despite the change in support and willingness to use, overall concern about the safety of these technologies very slightly decreased from 2018 to 2020. Overall, individuals were most concerned about the potential impact of autonomous weapon systems on civilians, with average safety concern scores of $1.572$ and $1.591$ in 2018 and 2020, respectively. (Note less concern indicates a higher mean, whereas more concern indicates a lower mean). In 2018, individuals also appeared more concerned about the safety of autonomous surgery, with a mean level of concern of $1.871$, which lessened to an average of $1.96$ in 2020, putting it more on par with the level of concern for militaries using autonomous weapon systems as well as autonomous vehicles, all which hovered around $1.851-1.718$. Respondents were least concerned about the safety of autonomous cyber defense, which remained constant at $1.96$.




\begin{figure}[H]
\centering
\includegraphics[width=0.8\textwidth]{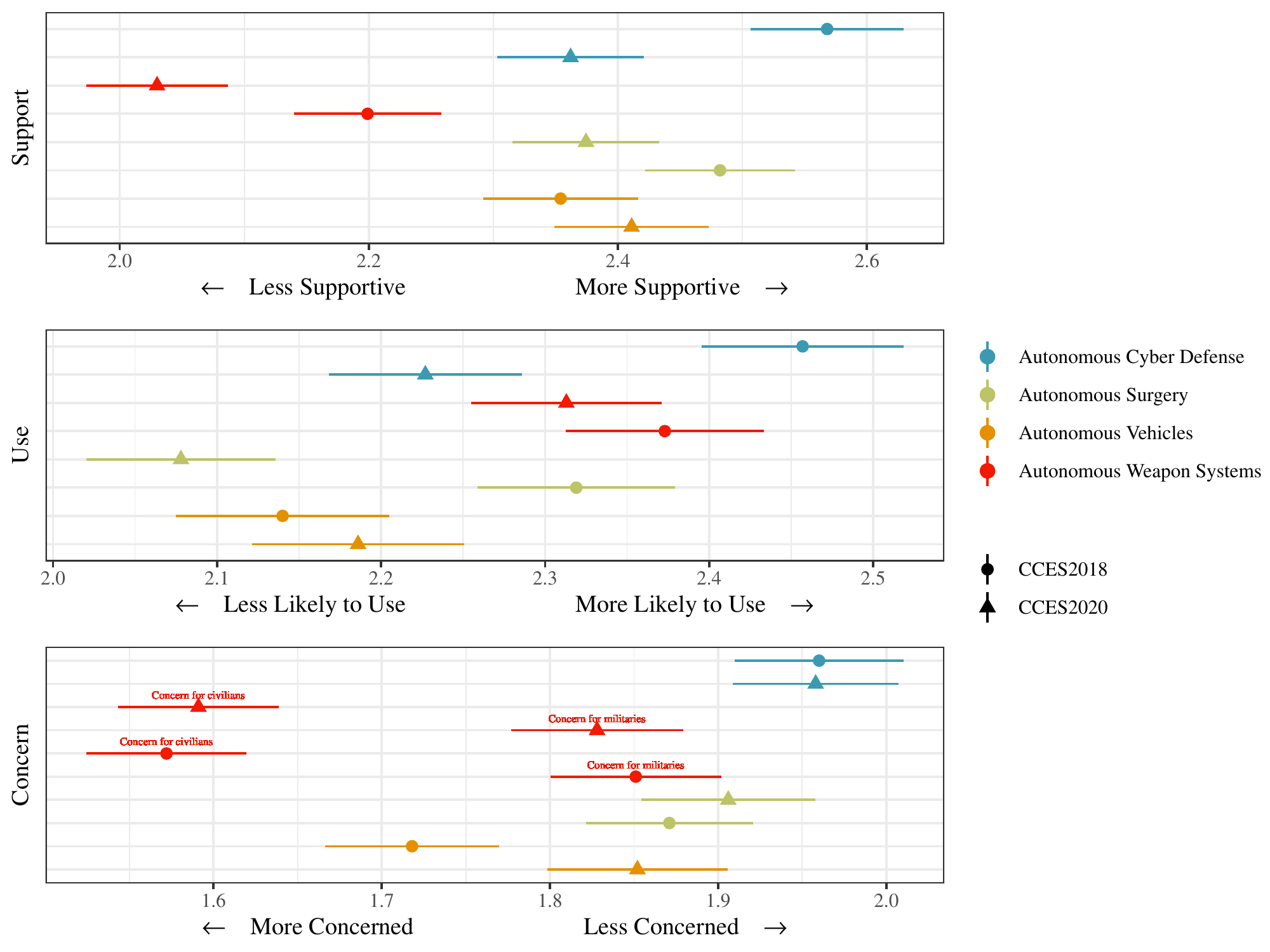}
\caption{Mean support, willingness to use, and concern for safety for all applications, 2018 vs 2020}
\label{fig:AllMeans}
\end{figure}

\subsection{Type of Application}

\begin{figure}[H]
\centering
\includegraphics[width=\textwidth]{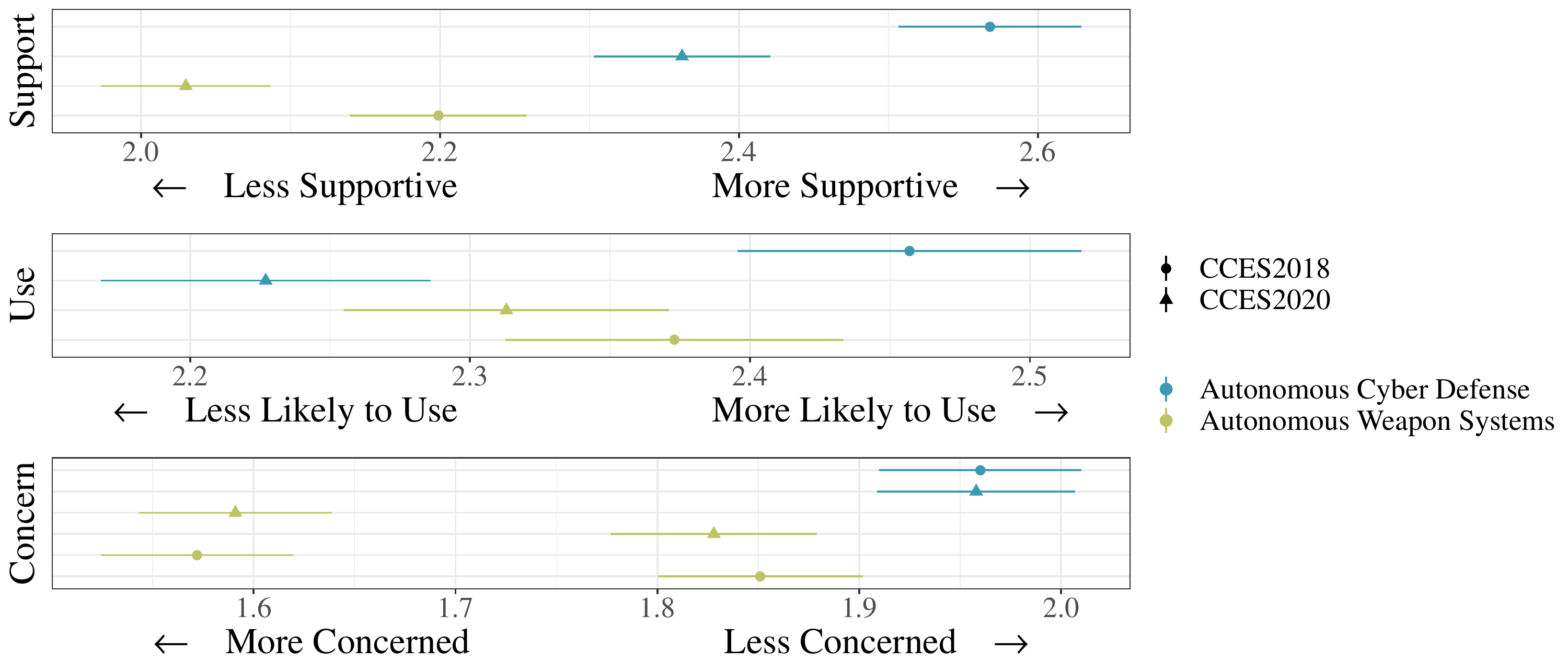}
\caption{Mean support, willingness to use, and concern for safety for autonomous cyber defense and autonomous weapons, 2018 vs 2020.}
\label{fig:CyberVsWeapons20182020Means}
\end{figure}

We also find support for hypothesis 3. The results confirm existing research \citep{young2018does, Horowitz2016} that autonomous weapon systems are controversial and face opposition from the general public. Figure \ref{fig:CyberVsWeapons20182020Means} below compares the support, use, and concern averages for autonomous weapon systems and another non-civilian use case --- AI cyber defense --- in both 2020 and 2018.

Why is support for autonomous weapon systems so low relative not only to civilian applications of AI but other potential military applications? One factor potentially at play is pop culture portrayals of AI-enabled weapons as dangerous in movies such as \textit{The Terminator}, and general ethical concerns. Since weapons inherently are potentially threatening, automating weapons comes with more substantial concerns of uncontrollable, dangerous technology.

The scenario in which support for autonomous weapon systems is the highest is when we ask respondents specifically about a situation of high importance for US national security. In that scenario, support for autonomous weapon systems rises to $2.31$, substantially higher than the $2.03$ average level of support for autonomous weapon systems in general. AI might contribute to higher rates of support under severe circumstances; AI weapons are perhaps seen as a necessity in a severe case and thereby could potentially justify setting aside ethical concerns.

\subsection{Prior experience and knowledge}

Hypothesis 1 focuses on how familiarity and prior experience with AI technologies and applications, measured via self-reported use and tested knowledge, should lead to increased positive sentiments including less concern, more support for, and a greater willingness to use AI-enabled autonomous systems.

\begin{table}[!htbp] \centering 
  \caption{Distribution of Self-Reported AI Knowledge} 
\begin{tabular}{@{\extracolsep{5pt}}lccccc} 
\\[-1.8ex]\hline\hline \\[-1.8ex] 
Response & \multicolumn{1}{c}{N}\\ 
\hline \\[-1.8ex] 
AI Index = 0            &     294& \\
AI Index = 1        &     354&      \\
AI Index = 2        &     203&       \\
AI Index = 3     &     103&  \\
AI Index = 4    &      43&   \\
AI Index = 5    &     3&   \\

\hline\hline \\[-1.8ex]
\end{tabular} 
  \label{SelfReportedAIKnowledge} 
\end{table} 

Most of the responses are clustered toward the lower end of the scale, with 65\% of the respondents reporting little to no prior use of AI. However, almost 30\% report a mid-level of prior experience/knowledge, with only a small number answering yes for all of the experience questions and answering one or both of the knowledge questions correctly. 

The results are broadly supportive of hypothesis 1---there is a statistically significant, positive relationship between the level of familiarity with AI and support for the adoption and use of autonomous systems for all applications except for autonomous weapon systems. The lack of a relationship between experience with AI and support for autonomous weapon systems is consistent with prior research on attitudes about LAWS from the general public and AI/ML experts \citep{zhang2021ethics, young2018does, Horowitz2016}. Tables \ref{tab:tablea1}-\ref{tab:tablea5} in the appendix and Figure \ref{fig:circlePlots} below display these results.

\begin{center}
\begin{figure} [H]
    \begin{subfigure}{.5\textwidth}
        \includegraphics[width=\linewidth, right]{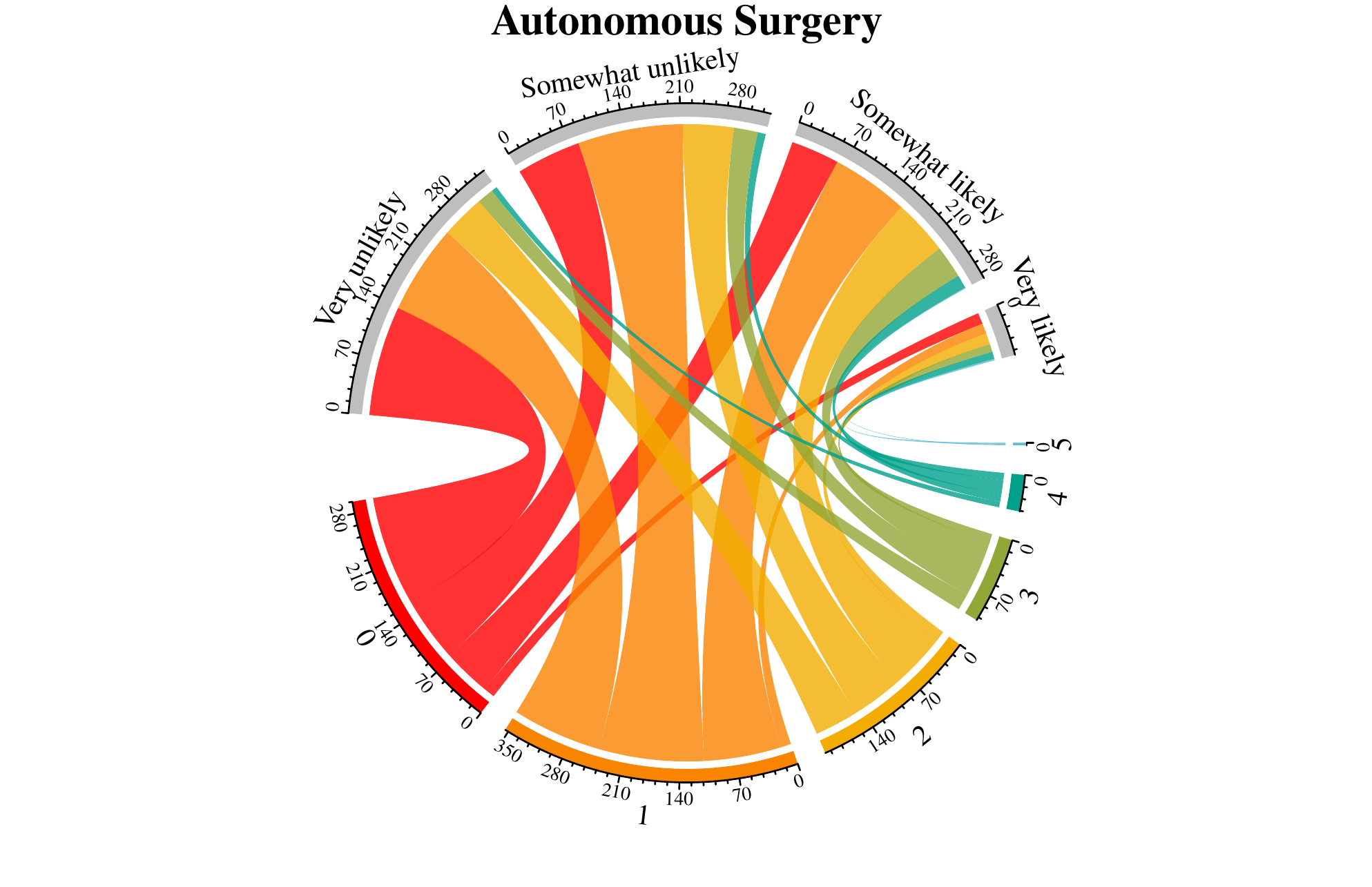}
    \end{subfigure}
    \hspace{.25cm}
    \begin{subfigure}{.5\textwidth}
            \includegraphics[width=\linewidth, left]{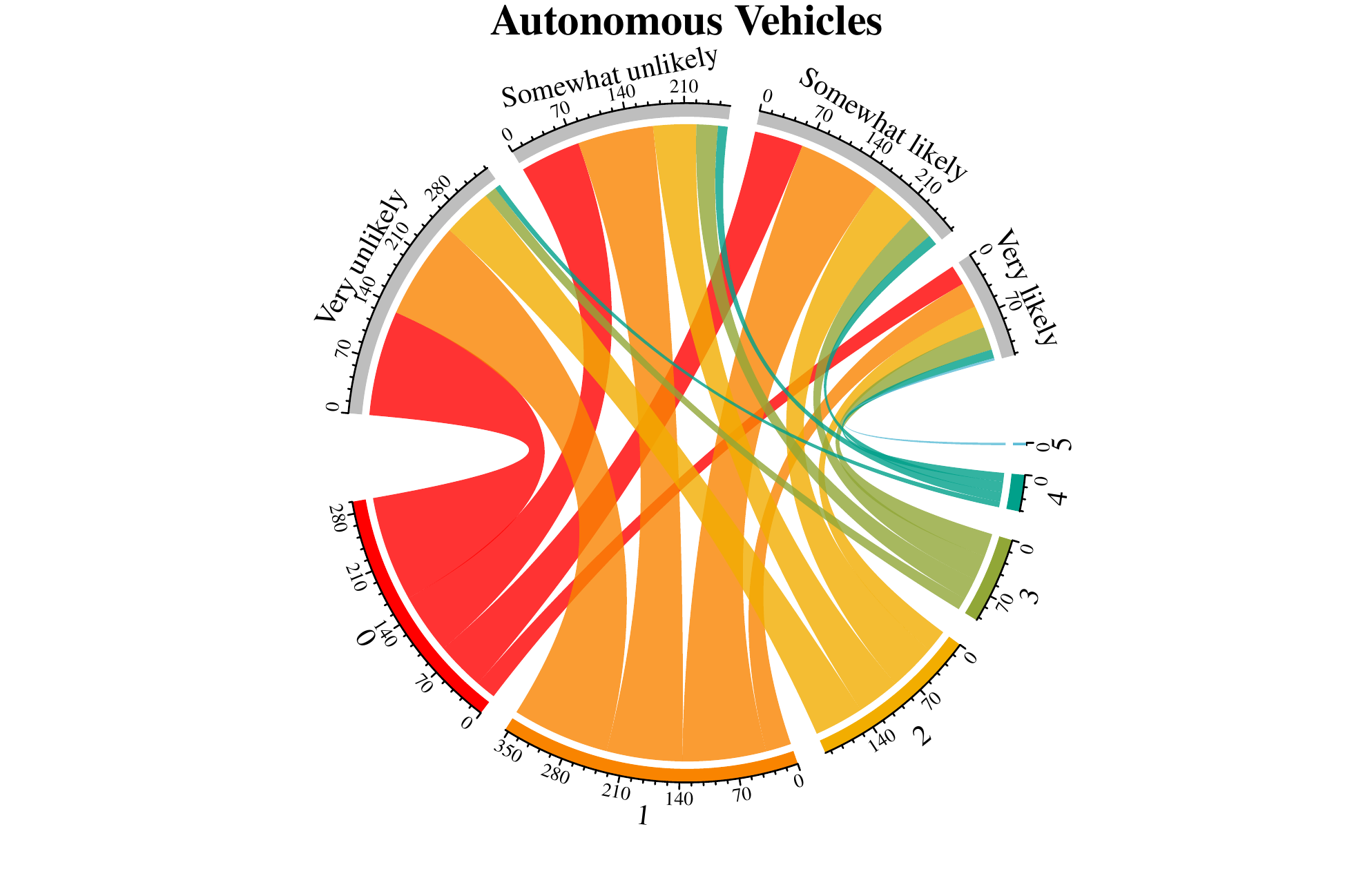}
    \end{subfigure} 
    \begin{subfigure}{.5\textwidth}
            \includegraphics[width=\linewidth, left]{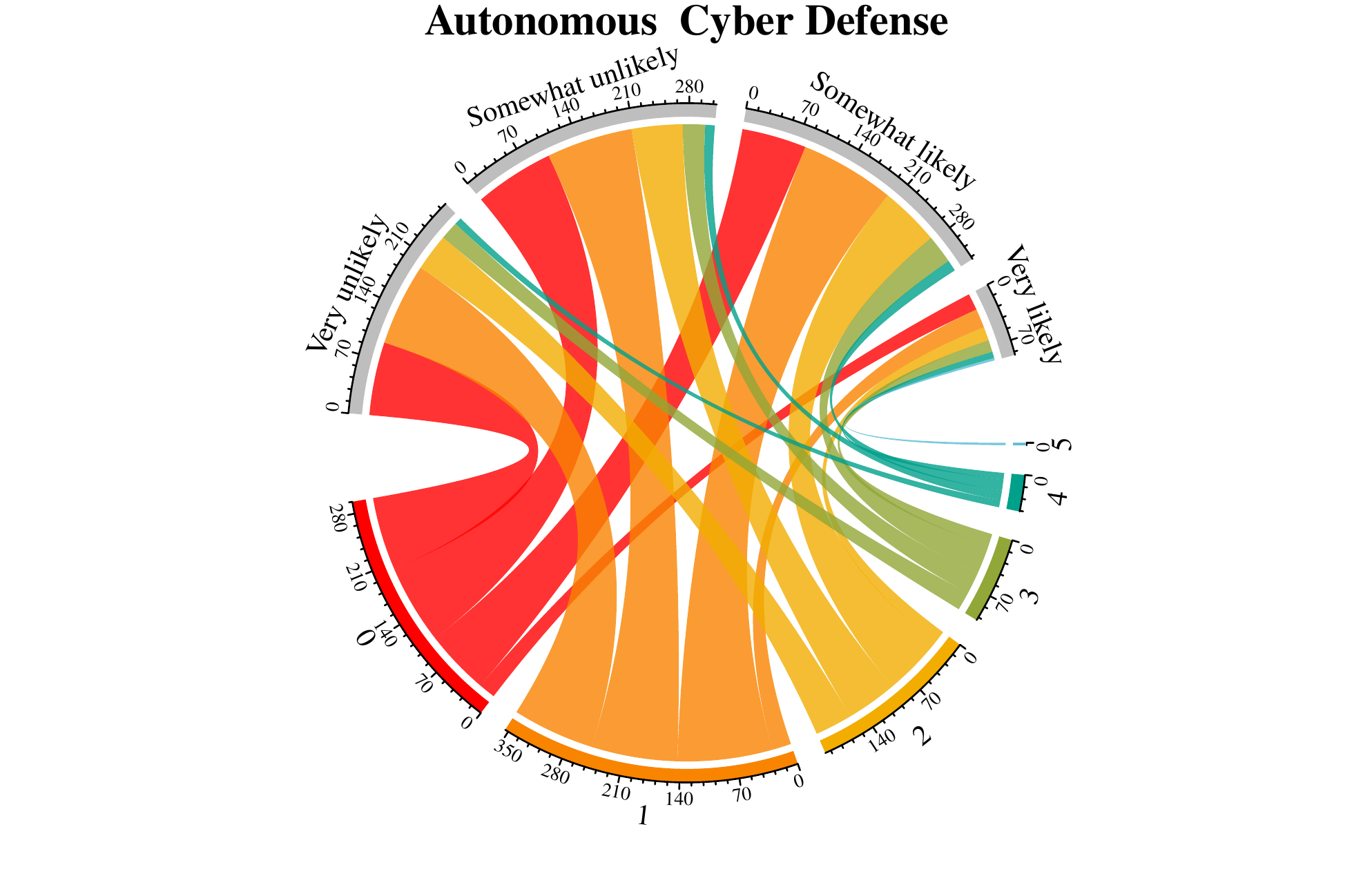}
    \end{subfigure}
    \hspace{.25cm}
    \begin{subfigure}{.5\textwidth}
            \includegraphics[width=\linewidth, left]{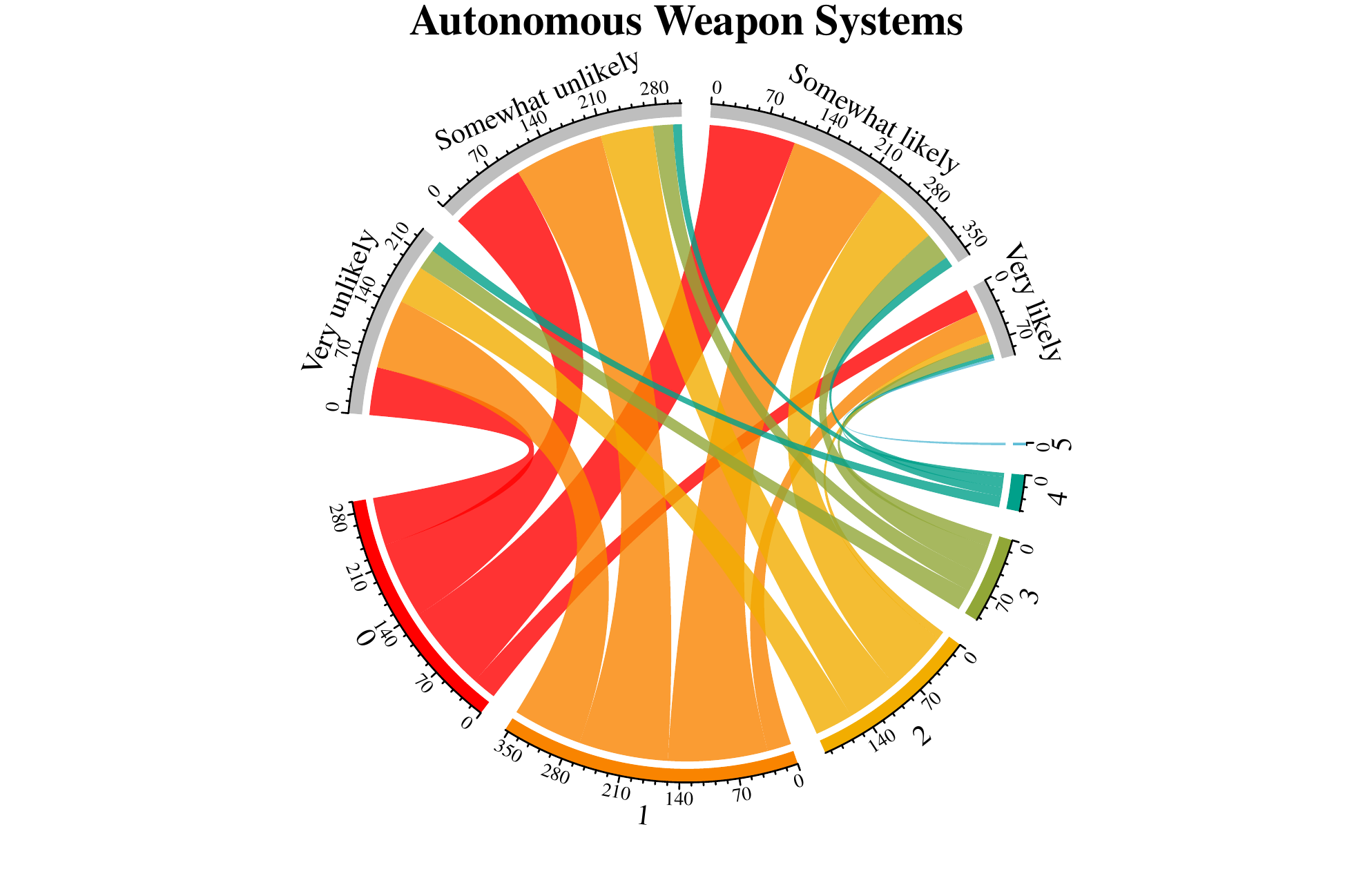}
    \end{subfigure}
    \caption{Willingness to personally use a given AI application, relative to the level of familiarity with AI (on a scale from 0 = no experience or knowledge of AI to 5 = has substantial machine learning knowledge and uses AI in multiple contexts.)}
    \label{fig:circlePlots} 
\end{figure}
\end{center}

\begin{figure}[H]
\centering
\includegraphics[width=0.8\textwidth]{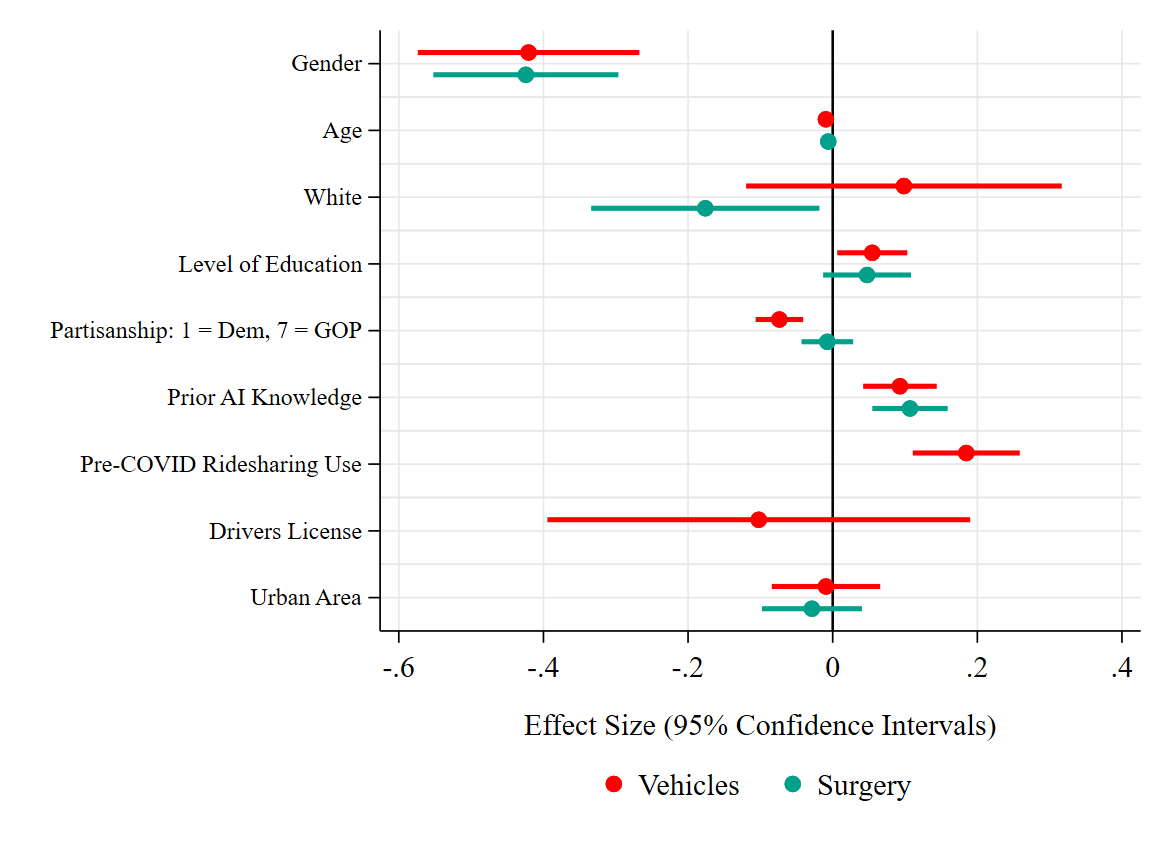}
\caption{Drivers of Support for Autonomous Vehicles and Surgery}
\label{fig:MarginsVehiclesSurgery}
\end{figure}

\begin{figure}[H]
\centering
\includegraphics[width=0.8\textwidth]{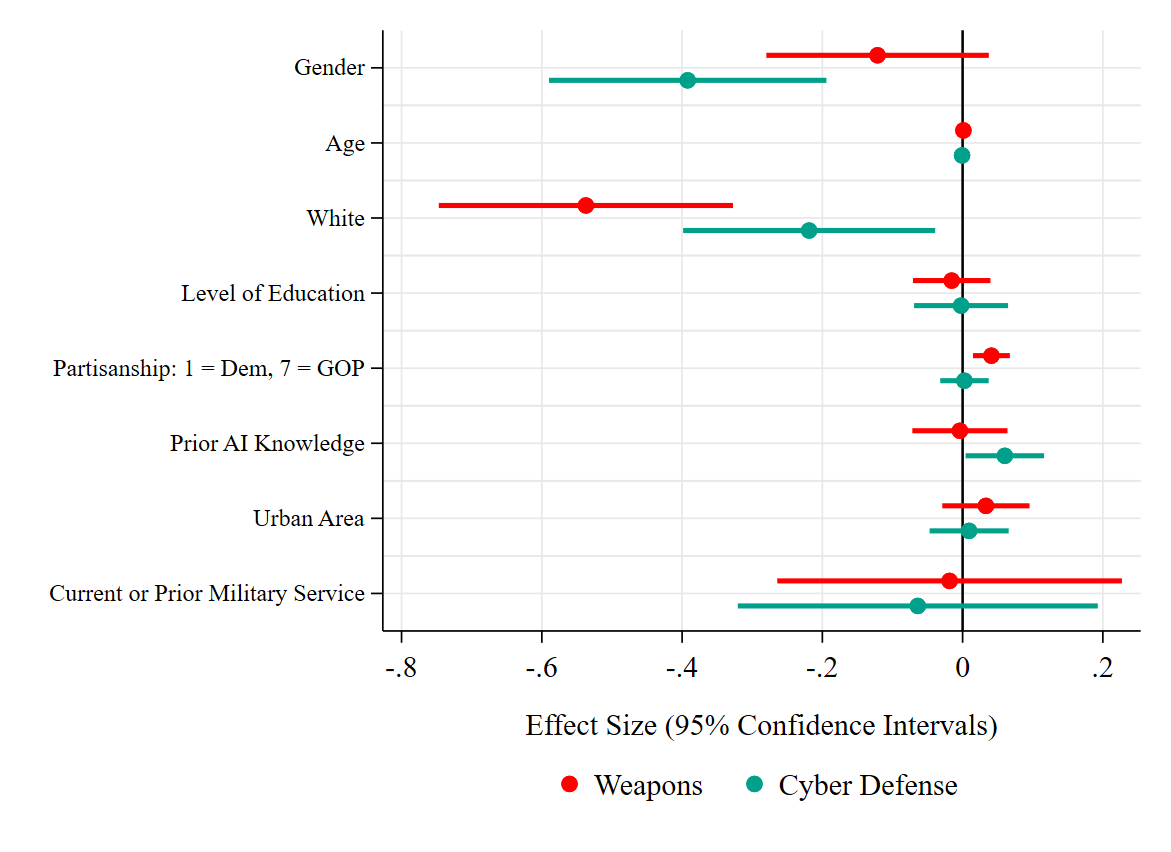}
\caption{Drivers of Support for Autonomous Weapon Systems and Cyber Defenses}
\label{fig:MarginsCyberAWS}
\end{figure}

To understand better the effects of prior knowledge and AI experience in context, we estimate OLS regression models to determine the relative effect of prior knowledge of AI and experience with AI on support for AI-enabled autonomous systems. The results, displayed in Figures \ref{fig:MarginsVehiclesSurgery} and \ref{fig:MarginsCyberAWS} below in the discussion, show generally strong substantive effects for prior AI knowledge. Moving from a low to a high level of prior AI knowledge generates a 9\% increase in support for autonomous vehicles, a 10\% increase in support for autonomous surgery, and a 6\% increase in support for autonomous cyber defense, with all of those increases statistically significant at the $0.05$ level or better. The lack of significance for the relationship between AI knowledge and experience and autonomous weapon systems is explained above.

To test which of the measures of AI knowledge and experience are driving the results, we re-run the main models shown in Figure \ref{fig:MarginsVehiclesSurgery} and \ref{fig:MarginsCyberAWS}, substituting in each of the components of the AI index in turn. The results, displayed in Tables \ref{tab:tablea1}-\ref{tab:tablea4} in the appendix, highlight how the experience variables are driving the results much more than the knowledge variables. Self-reported use of AI at home or work is positive and significantly associated ($p<0.05$) with support for autonomous vehicles, surgery, and weapons, and is positive but not significant for cyber defense. The use of AI to select music and movies is also positive and significantly associated ($p<0.05$) with support for autonomous vehicles, surgery, and cyber defense, but not autonomous weapon systems. Meanwhile, the AI knowledge questions were not statistically associated with greater support for AI adoption for any of the AI-enabled autonomous systems. What explains this result? One possibility is the knowledge questions, as displayed in the appendix, may have been too difficult. They asked respondents to identify what did and did not qualify as AI and machine learning, and that might have been too challenging. Future research should build on new attempts to test AI knowledge and awareness in the general public \citep{schepman2020}.

\subsection{Delegation}

We now evaluate our theory about delegation in the context of support for AI-enabled autonomous systems, especially autonomous vehicles, and surgery. We directly test this theory by looking at how those who used ridesharing apps prior to the COVID-19 pandemic feel about autonomous vehicles. Ridesharing users, after all, already made the decision to delegate driving to someone else, so they should be more supportive of self-driving cars than those that did not use ridesharing apps. 

\begin{table}[!htbp] \centering 
  \caption{Pre-COVID Ridesharing Use and Attitudes about Autonomous Vehicles} 
\begin{tabular}{@{\extracolsep{5pt}}lccccc} 
\\[-1.8ex]\hline\hline \\[-1.8ex] 
Response & \multicolumn{1}{c}{N} & \multicolumn{1}{c}{\shortstack{Percent\\That Support}} & \multicolumn{1}{c}{\shortstack{Percent\\Unconcerned}} & \multicolumn{1}{c}{\shortstack{Percent\\That Would Use}} \\ 
\hline \\[-1.8ex] 
Never            &     564& 42\% & 15\% & 32\% \\
A Few Times a Year       &     234&  63\% & 24\% & 50\%  \\
A Few Times a Month        &     89&   67\% & 24\% & 58\%   \\
A Few Times a Week     &     33& 64\% & 39\%  & 64\% \\
Almost Every Day or More    &      13&  62\% & 38\% & 85\% \\

\hline\hline \\[-1.8ex]
\end{tabular} 
  \label{ridesharing}
\end{table} 

In table \ref{ridesharing}, we show those somewhat or very supportive in each category as a percentage of the total number of respondents in that category of ridesharing users. 42\% of the 564 respondents that never used ridesharing were somewhat or very supportive of autonomous vehicles, but that percentage jumps above 60\% for all categories of respondents that used ridesharing prior to the COVID-19 pandemic. Similarly, those that used ridesharing are substantially less likely to be concerned about the safety of autonomous vehicles, and more likely to report they would personally use autonomous vehicles.

\begin{center}
\begin{figure} [H]
    \begin{subfigure}{.5\textwidth}
        \includegraphics[width=\linewidth, right]{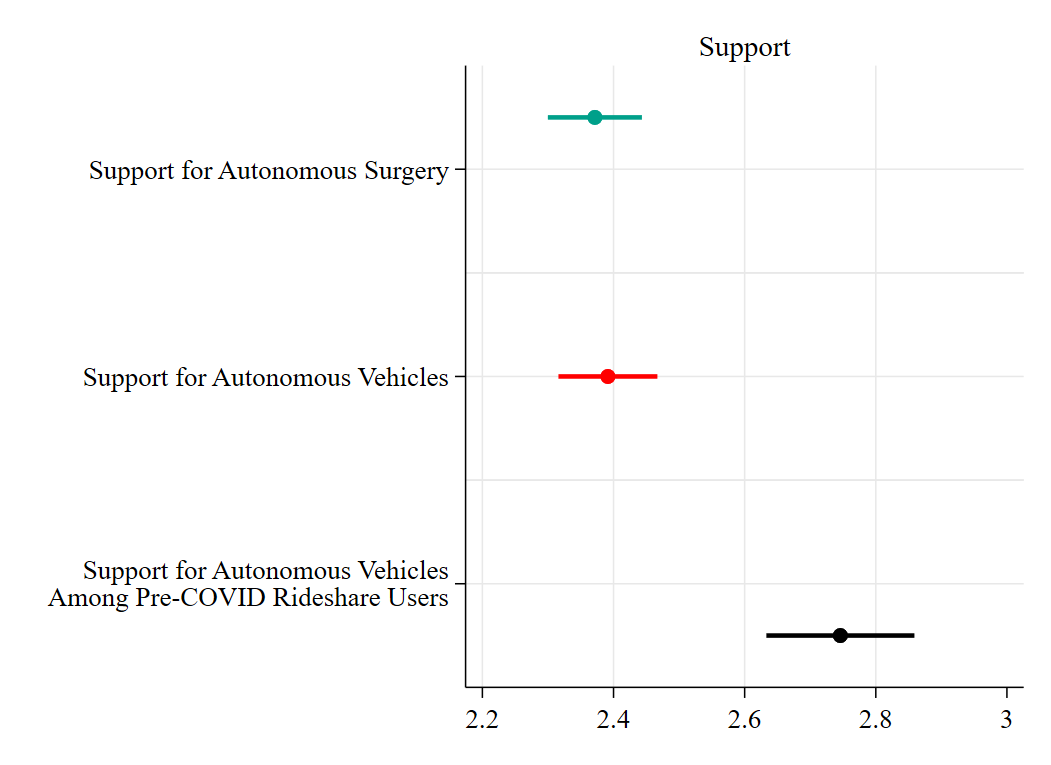}
    \end{subfigure}
    \hspace{.25cm}
    \begin{subfigure}{.5\textwidth}
            \includegraphics[width=\linewidth, left]{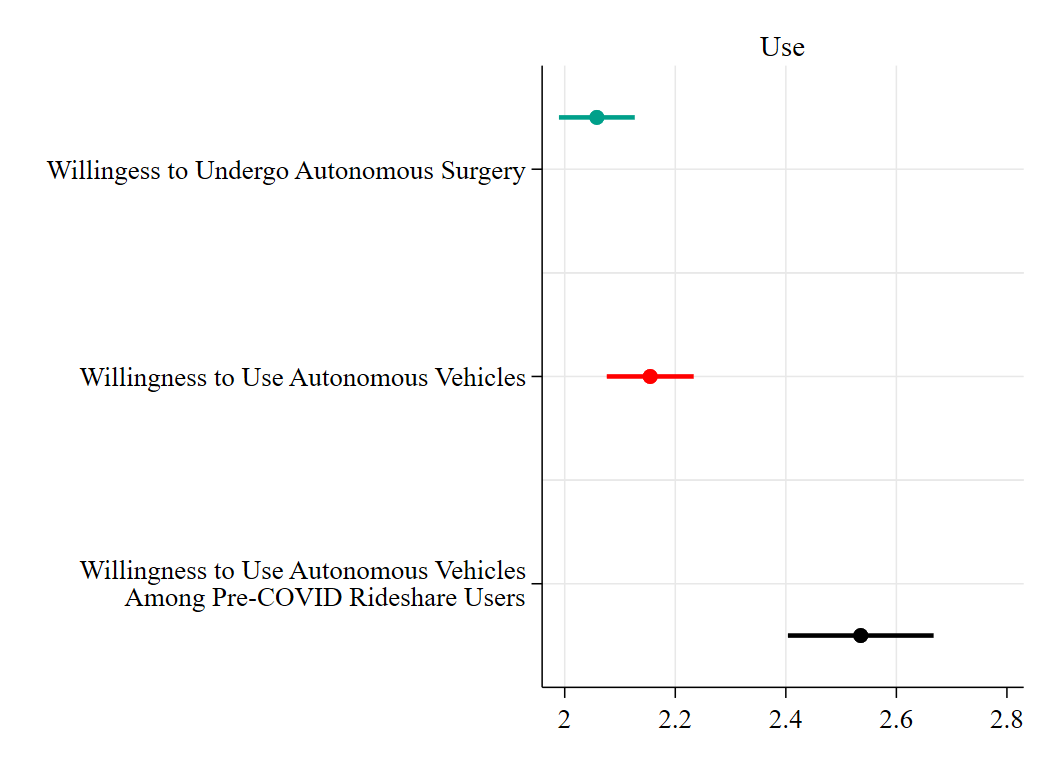}
    \end{subfigure} 
    \caption{Impact of Pre-COVID-19 Ridesharing Use on Willingness to Support or Use Autonomous Vehicles. Average and 95\% Confidence Interval}
    {\footnotesize Note: Higher numbers equal a higher degree of support}
    \label{fig:RelativeSupportVehiclesRidesharing} 
\end{figure}
\end{center}

This provides initial support for hypothesis 2a, which is reinforced in Figure \ref{fig:RelativeSupportVehiclesRidesharing}. Support for autonomous vehicles rises from an average of $2.39$ for all respondents to $2.75$ for those that used ridesharing pre-COVID. Similarly, personal willingness to use autonomous vehicles grows from an average of $2.15$ for all respondents to $2.54$ for those that used ridesharing pre-COVID. These gaps clearly show how those that delegated driving to others prior to the COVID-19 pandemic are more supportive of autonomous vehicles, as predicted.

\vspace{-1.5cm}

Support for hypothesis 2a is further confirmed when we shift to a regression context, based on the regression analysis described above in the context of hypothesis 1 and displayed in Figure \ref{fig:MarginsVehiclesSurgery}. Prior use of ridesharing apps has a large substantive effect---leading to an increase of almost 20\% in support for autonomous vehicles even when controlling for a range of demographic factors and prior AI knowledge and experience.

We also test for delegation effects for autonomous vehicles by looking at the sub-population of those in our sample that do not have a driver's license. By definition, they have already delegated driving to someone else. There are 123 respondents without a driver's license. They are more supportive of autonomous vehicles (average support $= 2.52$) than those with a driver's license (average support $= 2.37$), but the difference is not statistically significant at the $p<0.05$ level. The results for the personal use question are similar. The large confidence interval is likely driven by the small sample of non-drivers, so future research that over-samples on non-drivers could help address this issue. Alternatively, there might be health or mobility reasons why some people do not have a driver's license which might also limit the utility of autonomous vehicles for them, confounding any findings.

The results do not support hypothesis 2b concerning the relationship between autonomous surgery and vehicles. As Figure \ref{fig:RelativeSupportVehiclesRidesharing} shows, for the support and use questions, excluding those who used rideshare apps prior to COVID-19, there is no statistically significant difference between the averages for autonomous surgery and autonomous vehicles. In fact, for the use question, approval of the use of autonomous vehicles, even among those that did not use ridesharing apps, is higher than the approval of the use of autonomous surgery, though the difference is not statistically significant.

What explains the lack of a hypothesized result?  We can speculate. One explanation is that the type OF delegation is not the same across these use cases, since one involves the delegation of a “daily” activity, and one involves the delegation of a “rare” activity. Daily activities are things such as driving (even if everyone does not drive every day, driving for people with a driver's license is often commonplace, if not a frequent activity). Driving is a regular activity for most people, and though it is quite dangerous, given the number of accidents and accident-related deaths and injuries per year, it is probably perceived as less dangerous since it is familiar \citep{Guerin1994,ShariffEtAl2021}. For driving, delegation is a decision that can be adapted or changed dynamically, in the moment, based on circumstances and the comfort level of the individual doing the delegating or driving. Rare activities are those such as surgery, which is often also perceived as inherently dangerous. In a surgery case, the human cannot manage the risk themselves and is sometimes not awake or cognitively aware of the act of surgery itself. Additionally, whereas with driving, nearly all adults have experience as both a driver and a passenger, with surgery, unless you are a surgeon, you do not have that experience on the other side of the patient-surgeon interaction. Thus, with surgery, the decision to delegate is a forced choice, and an unfamiliar experience, and so not entirely parallel to that of autonomous vehicles. Similarly, individuals also cannot individually manage the risk themselves when it comes to national defense decisions.

Furthermore, one could argue it is not true delegation in the surgery and defense cases, as prior to any actual action being taken, the responsibility and procedures are already clearly established, with clear norms, guidelines, expectations, and requirements such as attending medical school or joining the military. As other research has highlighted, "when this division of labor is done by a designer prior to operation, it is a part of the design for that system," however, when this is done by a supervisor, human team, or individual dynamically during it, such as in the case of driving, "the process may be called 'delegation' or, more generally, 'tasking' and task management" \citep{Milleretal2007}. Thus, delegation as the theory section imagined is not appropriate, as delegation for a low-barrier, everyday activity such as driving is not comparable to infrequent activities that require specialized knowledge, membership, or access such as surgery or defense.  This is an important avenue for future research.

\subsection{Policy support vs. Use}

\begin{figure}[H]
\centering
\includegraphics[width=0.8\textwidth]{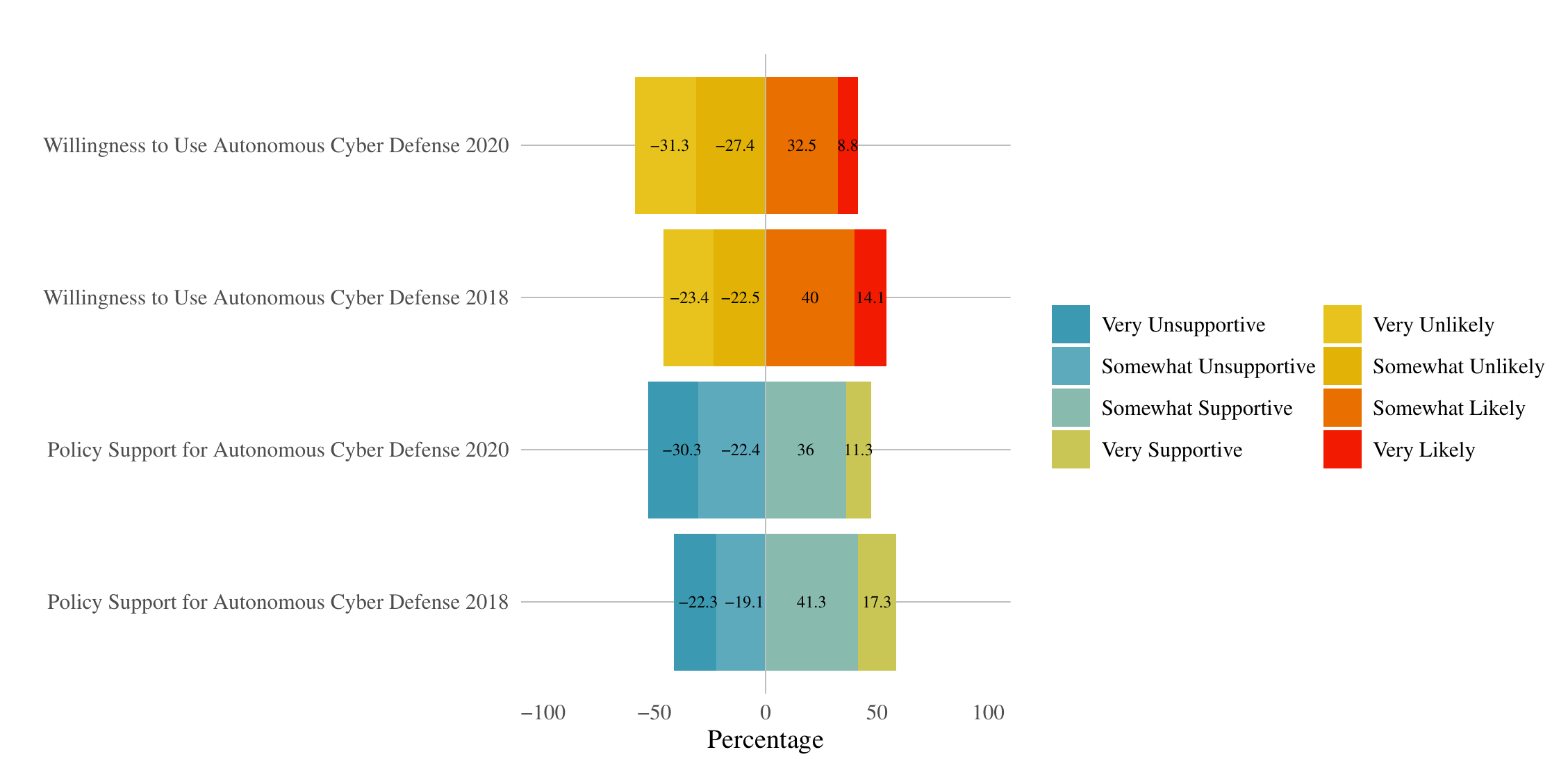}
\caption{Percentage support for/willingness to use autonomous cyber defense.}
\label{fig:supportUseGapCyber}
\end{figure}

\begin{figure}[H]
\centering
\includegraphics[width=0.8\textwidth]{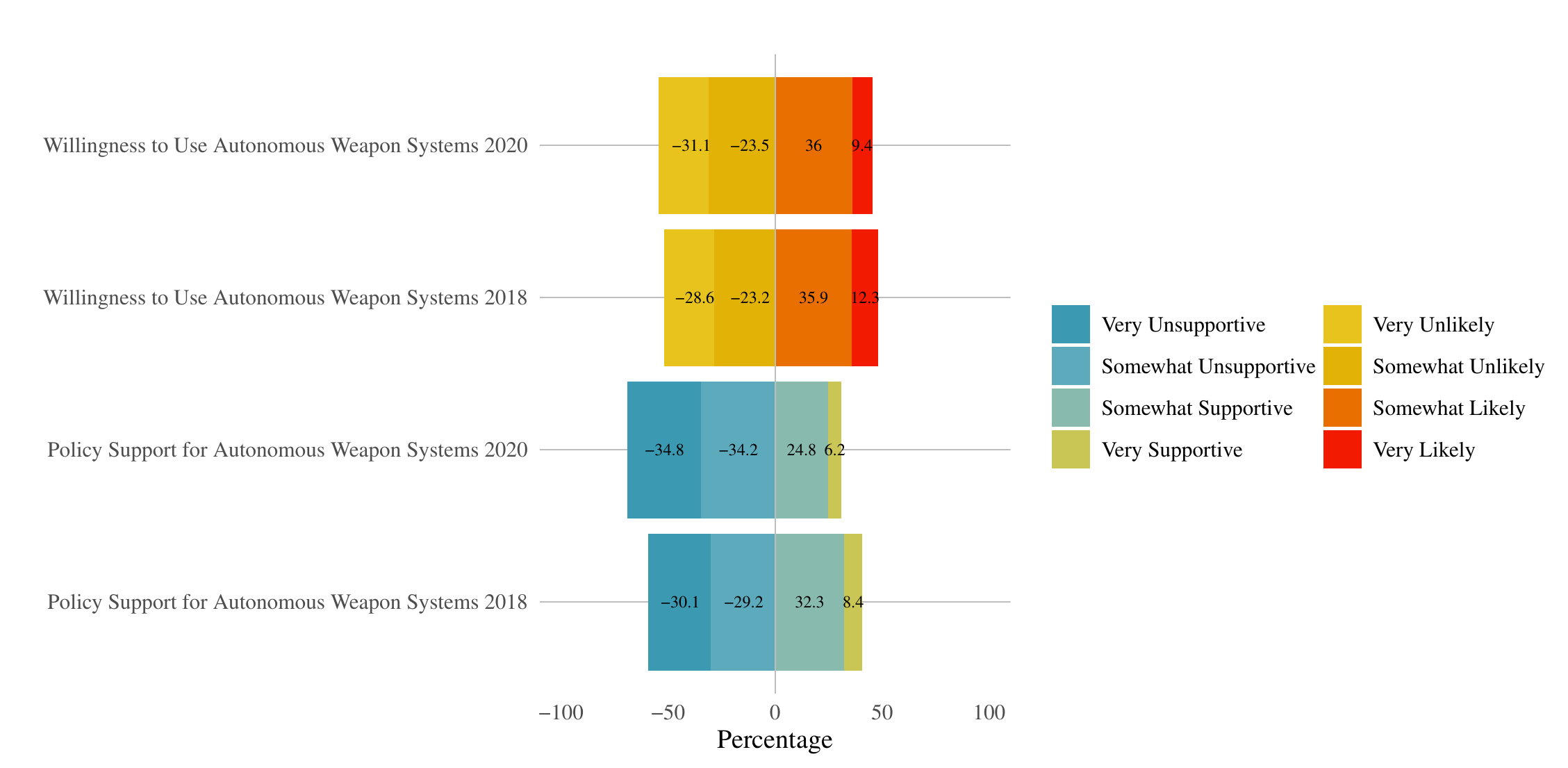}
\caption{Percentage support for/willingness to use autonomous weapon systems.}
\label{fig:supportUseGapAws}
\end{figure}

We now turn to assess whether people are more inclined to support these technologies in theory than to actually use them themselves. The results displayed below, in Figures \ref{fig:supportUseGapSurgery}, \ref{fig:supportUseGapVehicles}, \ref{fig:supportUseGapCyber}, \ref{fig:supportUseGapAws}, largely support hypothesis 5. There is a gap between support for development and willingness to use most of the autonomous systems we evaluate. Support for the systems as a matter of public policy is almost always higher than the willingness of individuals to use them.

Most respondents, on average, demonstrate a much lower personal willingness to undergo autonomous surgery than policy support generally. The same phenomenon is visible for vehicles, as well. This suggests there are some indicators that individuals might be excited by the broader societal benefits of these applications, but wary about the risks to the individual. However, when it came to military rather than civilian autonomous applications (weapons systems and cyber defense) the gap---between willingness to use and general support---diminished. Interestingly, for autonomous weapon systems in particular, on average, individuals were more encouraging of their "use to carry out a military mission of high importance to US national security" and shied away from supporting their general development more broadly.

\begin{figure}[H]
\centering
\includegraphics[width=0.8\textwidth]{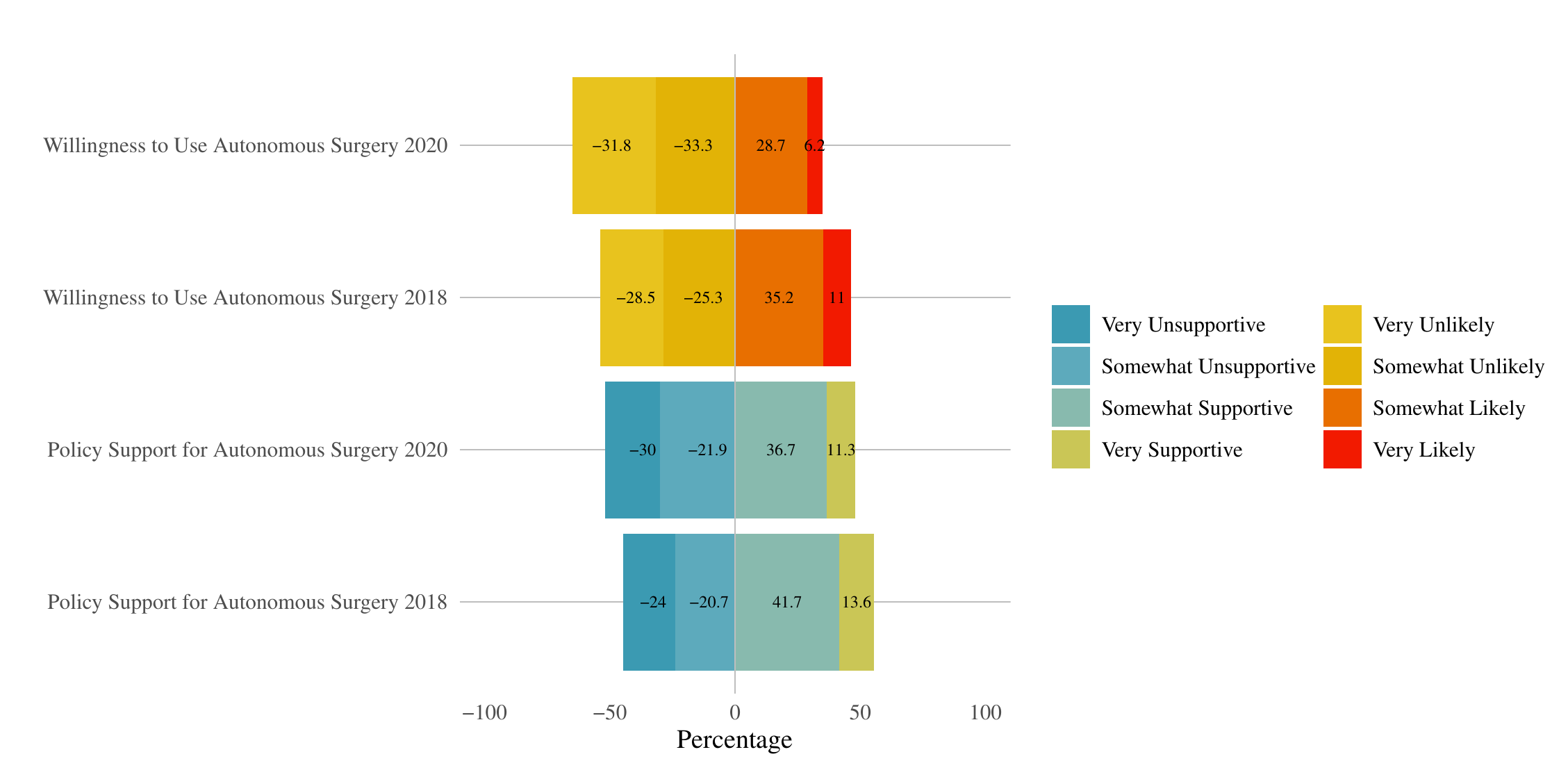}
\caption{Percentage support for/willingness to use autonomous surgery.}
\label{fig:supportUseGapSurgery}
\end{figure}

\begin{figure}[H]
\centering
\includegraphics[width=0.8\textwidth]{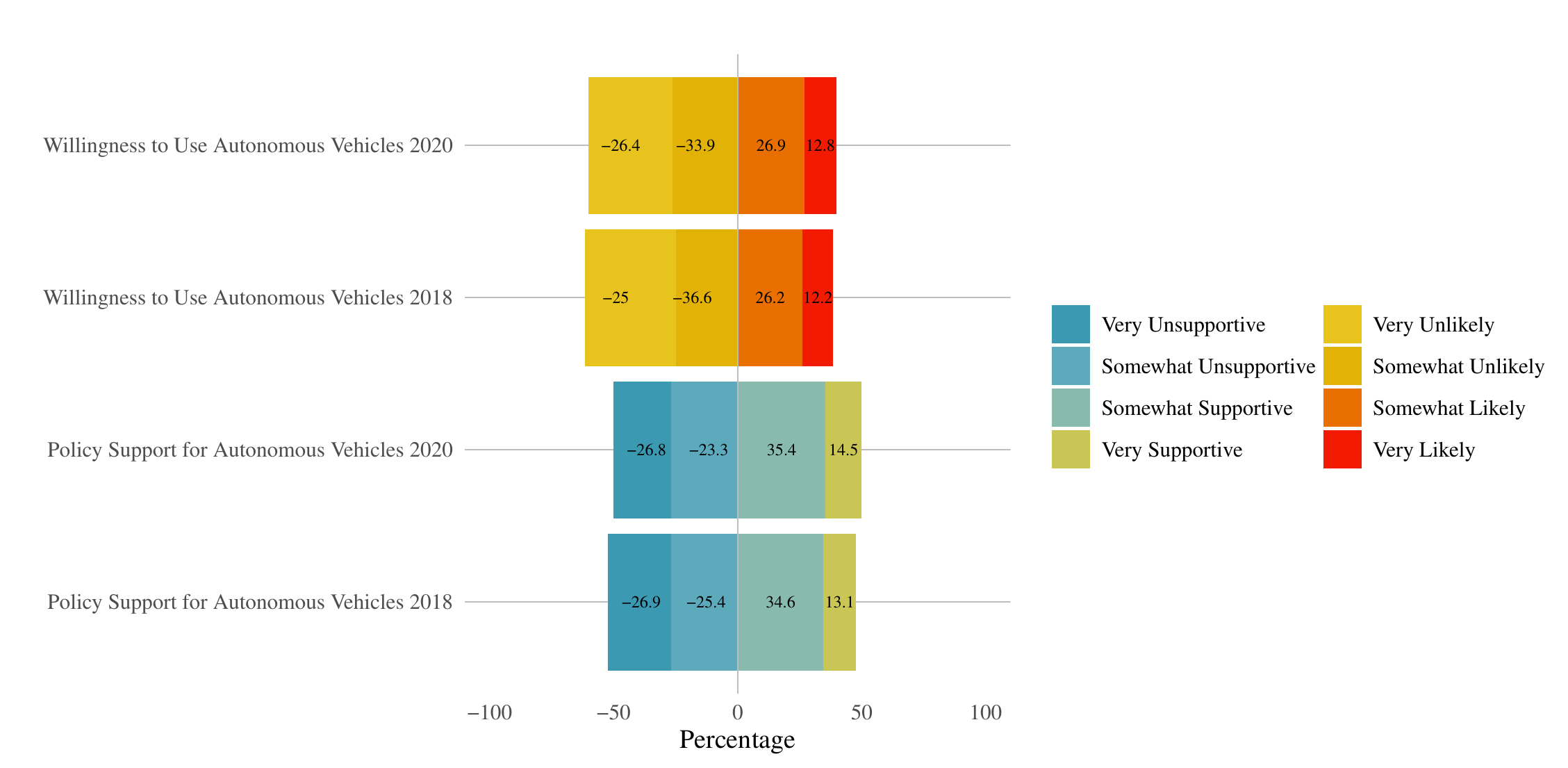}
\caption{Percentage support for/willingness to use autonomous vehicles.}
\label{fig:supportUseGapVehicles}
\end{figure}

\section{Demography and Ideology}

We now explore individual-level covariates and their relationship to support and use of these AI-enabled autonomous technologies. While we do not theorize about them, we include them as control variables. We discuss their importance, or lack thereof below, despite not hypothesizing about them, to lay the groundwork for future research and further contribute to the literature in more descriptive fashion. In general, research suggests emerging technologies are more likely to be adopted by younger, male, high-income individuals that work within technological fields \citep{KadylakCotten2020,MoodyEtAl2020,PayreEtAl2014,BansalEtAl2016,HabouchaEtAl2017,WangEtAl2020}. It's not a surprise this demographic is the most likely to use autonomous vehicles, especially as they also display a higher risk propensity for more general technology adoption \citep{HulseEtAl2018}. Moreover, attitudes about emerging technologies and science and technology issues are often polarized  \citep{gauchat2012, drummond2017, guber2013}. What do our results show? We now use OLS regression models, where the dependent variable is the level of support, and the independent variables are the covariates described in the Research Design section. We employ team weights to ensure population representation. The models are consistent---using ordered logit models, logit models, and without team weights.

The results show support for AI-enabled vehicles and surgery is substantially lower for women than men, with effect sizes that suggest a 40\% relative decline in support. Age is negative, but the substantive effects are very small, while higher levels of education, consistent with the literature, lead to stronger support for AI-enabled vehicles and surgery. Being in an urban area is not significantly associated with support for vehicles or surgery.

There are partisanship effects for vehicles, but not surgery. Republicans are significantly less likely, all else equal, to support autonomous vehicles, but there is no significant effect for autonomous surgery. Those that live in top 10 auto manufacturing states like Michigan are substantially less likely to support AI-enabled autonomous vehicles, with a 20\% drop in support---though the confidence interval is quite larger. Those in the top 10 healthcare employment states are more likely to support AI-enabled autonomous surgery.

The results for AI-enabled military systems differ in some ways from the vehicles and surgery results. There are strong gender effects for cyber defenses, with women less likely to support them than men, but while the coefficient is negative, there is not a statistically significant gender gap for autonomous weapon systems (perhaps because men are less likely to support them than any other AI-enabled autonomous system). There are no age effects, but there are race effects. Non-white respondents are substantially less likely to support autonomous weapon systems and autonomous cyber defenses, which we did not anticipate. This requires further investigation to understand why.

Higher levels of education, unlike for vehicles and surgery, do not lead to stronger support for autonomous weapon systems and cyber defenses. Prior military service is also not associated with stronger support. The only other clear effect comes from partisanship. Republicans are more likely to support autonomous weapon systems than Democrats (though not autonomous cyber defenses). This potentially reflects stronger Republican support, on average, for military systems.

\section{Conclusion}

This paper provides important new context for how familiarity with technology and previous delegation of related life decisions may influence the politics of support for AI-enabled autonomous systems. Across two representative surveys of US adults, we find those individuals with more experience using technology in the contexts of transportation and delegating the actual responsibility of driving via ridesharing apps prior to COVID-19 were more likely to support the adoption and use of AI-enabled autonomous systems in most cases. 

We also show individuals are more willing to support the development of these technologies than they are to actually use them themselves, suggesting while these technologies are interesting to the public, and the benefits they might provide, it is possible they are not yet familiar or convinced enough by the current state of the technology to use them comfortably in their daily lives. Finally, we also found any support, interest, or openness to these AI-enabled autonomous systems differs depending on use case. In particular, there exists a persistent, strong, aversion to autonomous weapon systems across key demographic categories.

There are limitations to these findings and the research design that could inform future research. We only survey US adult respondents. A larger, more global sample could further break down and show whether these perceptions of AI-enabled autonomous systems are general, or specific to certain contexts and cultures. Future research could also integrate more sub-populations to focus on how their views differ from those of the general public, such as healthcare providers’ view on autonomous surgery specifically, or how those who work in the military view autonomous cyber defense and weapon systems. Evaluating sub-populations with a close view of specific applications of AI-enabled autonomous systems will allow researchers to further explore how familiarity and the potential ability to perceive the benefits, risks, and uses of a given application alter support for their development. 

\section{Conflicts of Interest}

This article was made possible, in part, by a grant from the Air Force Office of Scientific Research and the Minerva Research Initiative under grant \#XXXXX-XX-X-XXXX. Number anonymized to preserve the anonymity of the authors in the review process. All content and errors are the responsibility of the authors. All data will be made available on a Dataverse page upon publication. The authors have no other competing interests to declare.

\newpage


\bibliographystyle{abbrvnat}
\bibliography{revisedpaper}

\newpage
\appendix

\counterwithin{figure}{section}
\counterwithin{table}{section}

\section{Online Appendix: Additional Tables and Figures}

\begin{table}[t!]
\centering 
\def\sym#1{\ifmmode^{#1}\else\(^{#1}\)\fi}
\hypertarget{tablea1}{}
\caption{Self-Reported AI Use and Support for Autonomous Vehicles}
\resizebox{.87\textwidth}{!}{
\begin{tabular}{l*{4}{c}}
\hline\hline
                    &\multicolumn{1}{c}{(1)}&\multicolumn{1}{c}{(2)}&\multicolumn{1}{c}{(3)}&\multicolumn{1}{c}{(4)}\\
                    &\multicolumn{1}{c}{\shortstack{Full AI\\Index}}&\multicolumn{1}{c}{\shortstack{AI Home\\Work}}&\multicolumn{1}{c}{\shortstack{AI Music\\Movies}}&\multicolumn{1}{c}{\shortstack{AI\\Knowledge}}\\
\hline
Prior AI Knowledge  &      0.0930\sym{***}&                     &                     &                     \\
                    &     (3.672)         &                     &                     &                     \\
Personal AI Use: 1 = home or work 2 = home/work&                     &       0.158\sym{***}&                     &                     \\
                    &                     &     (3.312)         &                     &                     \\
Personal AI Use: Music/Movies&                     &                     &       0.196\sym{**} &                     \\
                    &                     &                     &     (2.559)         &                     \\
AI Knowledge        &                     &                     &                     &      0.0193         \\
                    &                     &                     &                     &     (0.409)         \\
Gender              &      -0.421\sym{***}&      -0.432\sym{***}&      -0.439\sym{***}&      -0.447\sym{***}\\
                    &    (-5.515)         &    (-5.750)         &    (-5.501)         &    (-5.583)         \\
Age                 &    -0.00936\sym{***}&     -0.0102\sym{***}&    -0.00941\sym{***}&     -0.0109\sym{***}\\
                    &    (-4.477)         &    (-4.838)         &    (-4.710)         &    (-5.134)         \\
White               &      0.0985         &       0.115         &      0.0903         &      0.1000         \\
                    &     (0.907)         &     (1.101)         &     (0.827)         &     (0.936)         \\
Level of Education  &      0.0547\sym{**} &      0.0529\sym{**} &      0.0545\sym{**} &      0.0486\sym{**} \\
                    &     (2.268)         &     (2.190)         &     (2.280)         &     (2.050)         \\
Partisanship: 1 = Dem, 7 = GOP&     -0.0737\sym{***}&     -0.0757\sym{***}&     -0.0717\sym{***}&     -0.0769\sym{***}\\
                    &    (-4.507)         &    (-4.682)         &    (-4.326)         &    (-4.729)         \\
Pre-COVID Ridesharing Use&       0.185\sym{***}&       0.184\sym{***}&       0.194\sym{***}&       0.211\sym{***}\\
                    &     (5.019)         &     (5.093)         &     (5.051)         &     (5.618)         \\
Drivers License     &      -0.102         &     -0.0888         &      -0.102         &     -0.0502         \\
                    &    (-0.703)         &    (-0.615)         &    (-0.721)         &    (-0.355)         \\
Urban Area          &    -0.00930         &    -0.00901         &    -0.00842         &    -0.00980         \\
                    &    (-0.249)         &    (-0.239)         &    (-0.233)         &    (-0.269)         \\
Constant            &       2.820\sym{***}&       2.921\sym{***}&       2.854\sym{***}&       2.968\sym{***}\\
                    &    (14.065)         &    (14.622)         &    (13.752)         &    (15.472)         \\
\hline
Observations        &         886         &         886         &         886         &         886         \\
\(R^{2}\)           &       0.203         &       0.202         &       0.202         &       0.194         \\
Log Likelihood      &     -1155.1         &     -1155.6         &     -1155.8         &     -1160.0         \\
F                   &       19.65         &       16.52         &       19.32         &       17.87         \\

\hline\hline
\multicolumn{4}{l}{\footnotesize Notes: Standard errors clustered by state in parentheses. *p<0.10; **p< 0.05; ***p<0.01.}\\
\multicolumn{4}{l}{\footnotesize Notes: Observations weighted for representativeness}\\
\end{tabular}}
\label{tab:tablea1}
\end{table}

\begin{table}[H]
\centering 
\def\sym#1{\ifmmode^{#1}\else\(^{#1}\)\fi}
\hypertarget{tablea2}{}
\caption{Self-Reported AI Use and Support for Autonomous Surgery}
\resizebox{.87\textwidth}{!}{
\begin{tabular}{l*{4}{c}}
\hline\hline
                    &\multicolumn{1}{c}{(1)}&\multicolumn{1}{c}{(2)}&\multicolumn{1}{c}{(3)}&\multicolumn{1}{c}{(4)}\\
                    &\multicolumn{1}{c}{\shortstack{Full AI\\Index}}&\multicolumn{1}{c}{\shortstack{AI Home\\Work}}&\multicolumn{1}{c}{\shortstack{AI Music\\Movies}}&\multicolumn{1}{c}{\shortstack{AI\\Knowledge}}\\
\hline
Prior AI Knowledge  &       0.100\sym{***}&                     &                     &                     \\
                    &     (0.030)         &                     &                     &                     \\
Personal AI Use: 1 = home or work 2 = home/work&                     &       0.200\sym{***}&                     &                     \\
                    &                     &     (0.048)         &                     &                     \\
Personal AI Use: Music/Movies&                     &                     &       0.200\sym{***}&                     \\
                    &                     &                     &     (0.072)         &                     \\
AI Knowledge        &                     &                     &                     &      -0.004         \\
                    &                     &                     &                     &     (0.071)         \\
Gender              &      -0.455\sym{***}&      -0.464\sym{***}&      -0.475\sym{***}&      -0.487\sym{***}\\
                    &     (0.073)         &     (0.069)         &     (0.073)         &     (0.076)         \\
Age                 &      -0.006\sym{***}&      -0.007\sym{***}&      -0.006\sym{**} &      -0.008\sym{***}\\
                    &     (0.002)         &     (0.002)         &     (0.002)         &     (0.002)         \\
White               &      -0.067         &      -0.047         &      -0.076         &      -0.064         \\
                    &     (0.074)         &     (0.072)         &     (0.075)         &     (0.074)         \\
Level of Education  &       0.051\sym{**} &       0.050\sym{**} &       0.051\sym{**} &       0.044\sym{*}  \\
                    &     (0.024)         &     (0.024)         &     (0.025)         &     (0.025)         \\
Partisanship: 1 = Dem, 7 = GOP&      -0.019         &      -0.021         &      -0.017         &      -0.023         \\
                    &     (0.018)         &     (0.018)         &     (0.018)         &     (0.018)         \\
Pre-COVID Ridesharing Use&       0.089\sym{**} &       0.083\sym{**} &       0.100\sym{**} &       0.118\sym{***}\\
                    &     (0.038)         &     (0.038)         &     (0.038)         &     (0.037)         \\
Drivers License     &      -0.066         &      -0.059         &      -0.062         &      -0.009         \\
                    &     (0.131)         &     (0.126)         &     (0.129)         &     (0.122)         \\
Urban Area          &      -0.033         &      -0.032         &      -0.032         &      -0.033         \\
                    &     (0.030)         &     (0.030)         &     (0.030)         &     (0.029)         \\
Constant            &       2.715\sym{***}&       2.811\sym{***}&       2.758\sym{***}&       2.891\sym{***}\\
                    &     (0.265)         &     (0.278)         &     (0.261)         &     (0.267)         \\
\hline
Observations        &         886         &         886         &         886         &         886         \\
\(R^{2}\)           &       0.142         &       0.145         &       0.139         &       0.131         \\
Log Likelihood      &   -1148.885         &   -1147.332         &   -1150.257         &   -1154.731         \\
F                   &      14.022         &      17.181         &      14.185         &      14.223         \\

\hline\hline
\multicolumn{4}{l}{\footnotesize Notes: Standard errors clustered by state in parentheses. *p<0.10; **p< 0.05; ***p<0.01.}\\
\multicolumn{4}{l}{\footnotesize Notes: Observations weighted for representativeness}\\
\end{tabular}}
\label{tab:tablea2}
\end{table}

\begin{table}[H]
\centering 
\def\sym#1{\ifmmode^{#1}\else\(^{#1}\)\fi}
\hypertarget{tablea3}{}
\caption{Self-Reported AI Use and Support for Autonomous Weapon Systems}
\resizebox{.87\textwidth}{!}{
\begin{tabular}{l*{4}{c}}
\hline\hline
                    &\multicolumn{1}{c}{(1)}&\multicolumn{1}{c}{(2)}&\multicolumn{1}{c}{(3)}&\multicolumn{1}{c}{(4)}\\
                    &\multicolumn{1}{c}{\shortstack{Full AI\\Index}}&\multicolumn{1}{c}{\shortstack{AI Home\\Work}}&\multicolumn{1}{c}{\shortstack{AI Music\\Movies}}&\multicolumn{1}{c}{\shortstack{AI\\Knowledge}}\\
\hline
Prior AI Knowledge  &      -0.012         &                     &                     &                     \\
                    &     (0.029)         &                     &                     &                     \\
Personal AI Use: 1 = home or work 2 = home/work&                     &       0.088\sym{*}  &                     &                     \\
                    &                     &     (0.050)         &                     &                     \\
Personal AI Use: Music/Movies&                     &                     &       0.024         &                     \\
                    &                     &                     &     (0.074)         &                     \\
AI Knowledge        &                     &                     &                     &      -0.143\sym{**} \\
                    &                     &                     &                     &     (0.054)         \\
Gender              &      -0.126\sym{*}  &      -0.112         &      -0.120\sym{*}  &      -0.142\sym{*}  \\
                    &     (0.072)         &     (0.069)         &     (0.071)         &     (0.075)         \\
Age                 &       0.003         &       0.003         &       0.003         &       0.002         \\
                    &     (0.002)         &     (0.002)         &     (0.002)         &     (0.002)         \\
White               &      -0.460\sym{***}&      -0.452\sym{***}&      -0.461\sym{***}&      -0.451\sym{***}\\
                    &     (0.107)         &     (0.108)         &     (0.107)         &     (0.106)         \\
Level of Education  &      -0.010         &      -0.006         &      -0.008         &      -0.010         \\
                    &     (0.027)         &     (0.027)         &     (0.027)         &     (0.027)         \\
Partisanship: 1 = Dem, 7 = GOP&       0.030\sym{**} &       0.032\sym{**} &       0.031\sym{**} &       0.031\sym{**} \\
                    &     (0.013)         &     (0.013)         &     (0.013)         &     (0.013)         \\
Pre-COVID Ridesharing Use&       0.146\sym{***}&       0.127\sym{***}&       0.140\sym{***}&       0.146\sym{***}\\
                    &     (0.039)         &     (0.037)         &     (0.040)         &     (0.039)         \\
Drivers License     &      -0.168         &      -0.197\sym{*}  &      -0.181\sym{*}  &      -0.167\sym{*}  \\
                    &     (0.105)         &     (0.102)         &     (0.105)         &     (0.100)         \\
Urban Area          &       0.026         &       0.027         &       0.026         &       0.027         \\
                    &     (0.025)         &     (0.025)         &     (0.025)         &     (0.025)         \\
Constant            &       2.131\sym{***}&       2.076\sym{***}&       2.094\sym{***}&       2.210\sym{***}\\
                    &     (0.178)         &     (0.173)         &     (0.164)         &     (0.186)         \\
\hline
Observations        &         886         &         886         &         886         &         886         \\
\(R^{2}\)           &       0.075         &       0.078         &       0.075         &       0.083         \\
Log Likelihood      &   -1150.363         &   -1149.021         &   -1150.390         &   -1146.662         \\
F                   &      20.998         &      20.500         &      19.909         &      21.262         \\

\hline\hline
\multicolumn{4}{l}{\footnotesize Notes: Standard errors clustered by state in parentheses. *p<0.10; **p< 0.05; ***p<0.01.}\\
\multicolumn{4}{l}{\footnotesize Notes: Observations weighted for representativeness}\\
\end{tabular}}
\label{tab:tablea3}
\end{table}

\begin{table}[H]
\centering 
\def\sym#1{\ifmmode^{#1}\else\(^{#1}\)\fi}
\hypertarget{tablea4}{}
\caption{Self-Reported AI Use and Support for Autonomous Cyber Defense}
\resizebox{.87\textwidth}{!}{
\begin{tabular}{l*{4}{c}}
\hline\hline
                    &\multicolumn{1}{c}{(1)}&\multicolumn{1}{c}{(2)}&\multicolumn{1}{c}{(3)}&\multicolumn{1}{c}{(4)}\\
                    &\multicolumn{1}{c}{\shortstack{Full AI\\Index}}&\multicolumn{1}{c}{\shortstack{AI Home\\Work}}&\multicolumn{1}{c}{\shortstack{AI Music\\Movies}}&\multicolumn{1}{c}{\shortstack{AI\\Knowledge}}\\
\hline
Prior AI Knowledge  &       0.061\sym{**} &                     &                     &                     \\
                    &     (0.025)         &                     &                     &                     \\
Personal AI Use: 1 = home or work 2 = home/work&                     &       0.080         &                     &                     \\
                    &                     &     (0.053)         &                     &                     \\
Personal AI Use: Music/Movies&                     &                     &       0.121\sym{*}  &                     \\
                    &                     &                     &     (0.061)         &                     \\
AI Knowledge        &                     &                     &                     &       0.042         \\
                    &                     &                     &                     &     (0.054)         \\
Gender              &      -0.365\sym{***}&      -0.375\sym{***}&      -0.377\sym{***}&      -0.377\sym{***}\\
                    &     (0.090)         &     (0.089)         &     (0.089)         &     (0.091)         \\
Age                 &       0.000         &      -0.001         &       0.000         &      -0.001         \\
                    &     (0.002)         &     (0.002)         &     (0.002)         &     (0.002)         \\
White               &      -0.129         &      -0.120         &      -0.134         &      -0.130         \\
                    &     (0.098)         &     (0.098)         &     (0.100)         &     (0.098)         \\
Level of Education  &       0.008         &       0.006         &       0.007         &       0.004         \\
                    &     (0.032)         &     (0.032)         &     (0.031)         &     (0.032)         \\
Partisanship: 1 = Dem, 7 = GOP&      -0.006         &      -0.008         &      -0.005         &      -0.008         \\
                    &     (0.018)         &     (0.018)         &     (0.018)         &     (0.018)         \\
Pre-COVID Ridesharing Use&       0.094\sym{*}  &       0.098\sym{*}  &       0.101\sym{*}  &       0.111\sym{**} \\
                    &     (0.054)         &     (0.053)         &     (0.054)         &     (0.052)         \\
Drivers License     &      -0.160         &      -0.145         &      -0.157         &      -0.127         \\
                    &     (0.120)         &     (0.122)         &     (0.118)         &     (0.119)         \\
Urban Area          &       0.015         &       0.015         &       0.015         &       0.014         \\
                    &     (0.029)         &     (0.028)         &     (0.029)         &     (0.028)         \\
Constant            &       2.557\sym{***}&       2.633\sym{***}&       2.585\sym{***}&       2.634\sym{***}\\
                    &     (0.257)         &     (0.256)         &     (0.253)         &     (0.255)         \\
\hline
Observations        &         886         &         886         &         886         &         886         \\
\(R^{2}\)           &       0.066         &       0.064         &       0.065         &       0.062         \\
Log Likelihood      &   -1183.501         &   -1184.460         &   -1184.032         &   -1185.262         \\
F                   &      11.857         &       9.602         &      10.619         &      11.211         \\

\hline\hline
\multicolumn{4}{l}{\footnotesize Notes: Standard errors clustered by state in parentheses. *p<0.10; **p< 0.05; ***p<0.01.}\\
\multicolumn{4}{l}{\footnotesize Notes: Observations weighted for representativeness}\\
\end{tabular}}
\label{tab:tablea4}
\end{table}

\begin{table}[H]
\centering 
\def\sym#1{\ifmmode^{#1}\else\(^{#1}\)\fi}
\hypertarget{tablea5}{}
\caption{Support for AI Conditional on Self-Reported AI Experience}
\resizebox{.87\textwidth}{!}{
\begin{tabular}{l*{4}{c}}
\hline\hline
                    &\multicolumn{1}{c}{(1)}&\multicolumn{1}{c}{(2)}&\multicolumn{1}{c}{(3)}&\multicolumn{1}{c}{(4)}\\
                    &\multicolumn{1}{c}{\shortstack{Autonomous\\Vehicles}}&\multicolumn{1}{c}{\shortstack{Autonomous\\Surgery}}&\multicolumn{1}{c}{\shortstack{Autonomous\\Weapon Systems}}&\multicolumn{1}{c}{\shortstack{Autonomous\\Cyber Defense}}\\
\hline
Gender              &      -0.430\sym{**} &      -0.400\sym{***}&      -0.192         &      -0.382\sym{**} \\
                    &     (0.174)         &     (0.128)         &     (0.146)         &     (0.156)         \\
Age                 &      -0.006         &      -0.005         &       0.003         &      -0.003         \\
                    &     (0.006)         &     (0.005)         &     (0.004)         &     (0.006)         \\
White               &      -0.167         &      -0.362\sym{**} &      -0.537\sym{***}&      -0.198         \\
                    &     (0.143)         &     (0.153)         &     (0.142)         &     (0.176)         \\
Level of Education  &       0.110\sym{**} &       0.157\sym{**} &       0.012         &       0.119\sym{*}  \\
                    &     (0.047)         &     (0.066)         &     (0.073)         &     (0.065)         \\
Partisanship: 1 = Dem, 7 = GOP&                     &                     &                     &                     \\
                    &                     &                     &                     &                     \\
Pre-COVID Ridesharing Use&       0.080         &       0.082         &       0.109         &       0.002         \\
                    &     (0.066)         &     (0.087)         &     (0.083)         &     (0.118)         \\
Drivers License     &       0.249         &       0.007         &      -0.618\sym{**} &      -0.500\sym{**} \\
                    &     (0.197)         &     (0.313)         &     (0.231)         &     (0.227)         \\
Urban Area          &      -0.074         &      -0.116         &      -0.112         &      -0.113         \\
                    &     (0.074)         &     (0.082)         &     (0.094)         &     (0.071)         \\
GOP: 1 = lean, 2 = not strong, 3 = strong&      -0.004         &       0.040         &       0.159         &       0.299\sym{***}\\
                    &     (0.062)         &     (0.048)         &     (0.097)         &     (0.070)         \\
Constant            &       2.788\sym{***}&       2.840\sym{***}&       2.903\sym{***}&       3.118\sym{***}\\
                    &     (0.299)         &     (0.515)         &     (0.315)         &     (0.465)         \\
\hline
Observations        &         142         &         141         &         142         &         142         \\
\(R^{2}\)           &       0.131         &       0.152         &       0.180         &       0.159         \\
Log Likelihood      &    -166.382         &    -176.837         &    -182.501         &    -182.159         \\
F                   &       2.712         &       7.197         &      16.940         &       6.194         \\

\hline\hline
\multicolumn{4}{l}{\footnotesize Notes: Standard errors clustered by state in parentheses. *p<0.10; **p< 0.05; ***p<0.01.}\\
\multicolumn{4}{l}{\footnotesize Notes: Observations weighted for representativeness}\\
\end{tabular}}
\label{tab:tablea5}
\end{table}

\section{Survey Text}

\singlespacing 

\noindent Vehicles1

\noindent Prior to the COVID‐19 pandemic, how frequently did you use ridesharing apps such as Lyft and Uber?

\begin{itemize}
    \item Never
    \item A few times a year
    \item A few times a month
    \item A few times a week
    \item Almost every day or more
    \item I do not know what ridesharing apps are
\end{itemize}

\noindent Vehicles2

\noindent Imagine that automotive technology has advanced to the point where self‐driving cars and trucks require little or no input from humans to operate. Would you support or oppose this development?

\begin{itemize}
    \item Strongly support 
    \item Somewhat support 
    \item Somewhat oppose
    \item Strongly oppose 
\end{itemize}

\noindent Vehicles3

\noindent How concerned are you about the safety of autonomous vehicles for those riding in them as well as for other vehicles, cyclists, and pedestrians?
\begin{itemize}
    \item Not at all concerned
    \item Slightly concerned
    \item Very concerned
    \item Extremely concerned
\end{itemize}

\noindent Vehicles4

\noindent How likely would you be to ride in an autonomous vehicle?
\begin{itemize}
    \item Very likely
    \item Somewhat likely
    \item Somewhat unlikely
    \item Very unlikely
\end{itemize}

\noindent Healthcare1 

\noindent Imagine that medical technology has advanced to the point where surgical procedures will be able to be performed by autonomous systems (systems trained by an algorithm) with little to no input from humans. Would you support or oppose this development?
\begin{itemize}
    \item Strongly support 
    \item Somewhat support 
    \item Somewhat oppose
    \item Strongly oppose 
\end{itemize}

\noindent Healthcare2

\noindent How concerned are you about the safety of procedures for patients who undergo surgery conducted by autonomous systems?
\begin{itemize}
    \item Not at all concerned
    \item Slightly concerned
    \item Very concerned
    \item Extremely concerned
\end{itemize}

\noindent Healthcare3

\noindent Would you support or oppose the use of artificial intelligence to make decisions about the allocation of limited healthcare resources?
\begin{itemize}
    \item Strongly support 
    \item Somewhat support 
    \item Somewhat oppose
    \item Strongly oppose 
\end{itemize}

\noindent Weapons1 

\noindent Imagine that military technology has advanced to the point where autonomous weapon systems that require no input from humans after activation will be able to target and fire weapons. How supportive would you be of this development? 
\begin{itemize}
    \item Strongly support 
    \item Somewhat support 
    \item Somewhat oppose
    \item Strongly oppose 
\end{itemize}

\noindent Weapons2

\noindent How concerned are you about the safety of autonomous weapon systems for the militaries that operate them?
\begin{itemize}
    \item Not at all concerned
    \item Slightly concerned
    \item Very concerned
    \item Extremely concerned
\end{itemize}

\noindent Weapons3

\noindent How concerned are you about the safety of autonomous weapon systems for civilians?
\begin{itemize}
    \item Not at all concerned
    \item Slightly concerned
    \item Very concerned
    \item Extremely concerned
\end{itemize}

\noindent Weapons4

\noindent How likely would you be to support the use of autonomous weapon systems to carry out a military mission of high importance to US national security?
\begin{itemize}
    \item Very likely
    \item Somewhat likely 
    \item Somewhat unlikely
    \item Very unlikely 
\end{itemize}

\noindent Cyber1

\noindent Imagine that cyber defense technology has advanced to the point where automated responses to cyber attacks require no input from humans to operate after activation. How supportive would you be of this development?
\begin{itemize}
    \item Strongly support 
    \item Somewhat support 
    \item Somewhat oppose
    \item Strongly oppose 
\end{itemize}

\noindent Cyber2

\noindent How concerned are you about the safety of autonomous cyber defense for those systems utilizing the technology as well as critical infrastructure dependent on these systems?
\begin{itemize}
    \item Not at all concerned
    \item Slightly concerned
    \item Very concerned
    \item Extremely concerned
\end{itemize}

\noindent Cyber3

\noindent How likely would you be to use autonomous cyber defense technology?
\begin{itemize}
    \item Strongly support 
    \item Somewhat support 
    \item Somewhat oppose
    \item Strongly oppose 
\end{itemize}

\noindent AIHome-Work (1 point for AI Index for each of home and work)

\noindent Generally speaking, do you use artificial intelligence at work or at home?
\begin{itemize}
    \item At home
    \item At work
    \item At work and at home
    \item Neither
\end{itemize}

\noindent AIMusic-Movies (1 point for AI Index if yes)

\noindent Do you use artificial intelligence‐based systems to select music or movies for your enjoyment (e.g. Pandora or Netflix)?
\begin{itemize}
    \item Yes
    \item No
\end{itemize}

\noindent ML Knowledge A (1 point for AI Index if correct)

\noindent Choose the option that is not correct regarding artificial intelligence,
\begin{itemize}
    \item Artificial intelligence includes techniques that allow systems to learn without being explicitly programmed
    \item Machine learning is not a type of AI 
    \item Artificial intelligence is often categorized into two types: general Artificial intelligence and narrow Artificial intelligence 
    \item Artificial intelligence is a software, machine, or computer that researchers think could eventually emulate the human mind
\end{itemize}

\noindent ML Knowledge B (1 point for AI Index if correct)

\noindent Which of the following is NOT an example of supervised learning?
\begin{itemize}
    \item Principal Component Analysis
    \item Decision Tree
    \item Linear Regression 
    \item Naive Bayesian
\end{itemize}

\end{document}